\documentclass[a4paper,11pt]{article}
\pdfoutput=1 \usepackage{jheppub_2} 
\usepackage{booktabs}
\usepackage[T1]{fontenc} 
\usepackage{graphicx, multirow,soul,url,amsmath,amsfonts,amssymb,mathrsfs,amsfonts}
\usepackage[all]{hypcap}
\usepackage{url}
\usepackage{bbm}
\usepackage{changepage}
\usepackage[section]{placeins}
\usepackage{cancel,ulem}
\usepackage{sidecap}
\usepackage{graphicx}
\usepackage{slashed}
\usepackage[dvipsnames]{xcolor}
\usepackage{multicol, blindtext}
\usepackage[section]{placeins}
\usepackage[version=3]{mhchem}
\usepackage{caption}
\expandafter\def\csname ver@subfig.sty\endcsname{}
\usepackage{svg}
\usepackage{lipsum}
\usepackage{hyperref}
\usepackage{footnote}
\usepackage{bbold}
\usepackage{amsbsy}
\usepackage[export]{adjustbox}
\usepackage{units}

\usepackage{float}
\usepackage{pifont}
\restylefloat{table}
\usepackage{lipsum}

\def\thefootnote{\fnsymbol{footnote}}
\allowdisplaybreaks

\newcommand{\be}{\begin{equation}}
\newcommand{\ee}{\end{equation}}
\newcommand{\bea}{\begin{eqnarray}}
\newcommand{\eea}{\end{eqnarray}}
\newcommand{\ba}{\begin{aligned}}
\newcommand{\ea}{\end{aligned}}

\newcommand{\bq}{\begin{eqnarray}}
\newcommand{\nq}{\end{eqnarray}}

\usepackage{subfig}

\title{\bf Gravitational waves from cosmic superstrings and gauge strings}
\author[a]{Danny Marfatia}
\affiliation[a]{Department of Physics and Astronomy, University of Hawaii at Manoa, Honolulu, HI 96822, USA}
\author[b]{and Ye-Ling Zhou}
\affiliation[b]{School of Fundamental Physics and Mathematical Sciences, Hangzhou Institute for Advanced
Study, UCAS, Hangzhou 310024, China}
\emailAdd{dmarf8@hawaii.edu}
\emailAdd{zhouyeling@ucas.ac.cn}

\abstract{
We perform a phenomenological comparison of the gravitational wave (GW) spectrum expected from cosmic gauge string networks and superstring networks comprised of multiple string types.
We show how violations of scaling behavior and the evolution of the number of relativistic degrees of freedom in the early Universe affect the GW spectrum.
We derive simple analytical expressions for the GW spectrum from superstrings and gauge strings that are valid for all frequencies relevant to pulsar timing arrays (PTAs) and laser interferometers.
We analyze the latest data from PTAs and show that
superstring networks are consistent with 32~nHz data from NANOGrav, but are excluded by 3.2~nHz data at $3\sigma$ unless the string coupling $g_s<0.2$ or the strings evolve in 
only about 10\% of the volume of the higher-dimensional space.
We also point out that while gauge string networks are excluded by NANOGrav-15 data at $3\sigma$, they are completely compatible with EPTA and PPTA data. Finally, we study correlations between GW signals at PTAs and laser interferometers.
}


\begin{document}

\thispagestyle{empty}
\def\thefootnote{\fnsymbol{footnote}}
\setcounter{footnote}{1}

\setcounter{page}{0}
\maketitle
\vspace{-1cm}
\flushbottom

\def\thefootnote{\arabic{footnote}}
\setcounter{footnote}{0}

\section{Introduction}

Cosmic strings are astronomically large linear accumulations of energy. They may arise as topological defects during the spontaneous breaking of a symmetry in an early Universe phase transition \cite{Kibble:1976sj}, or as microscopic fundamental strings of superstring theory stretched to cosmological sizes by cosmic expansion \cite{Jones:2002cv,Sarangi:2002yt,Dvali:2003zj, Copeland:2003bj,Jackson:2004zg}. Cosmic strings interact with each other to form a string network which loses energy via radiation and displays a scaling behavior~\cite{Albrecht:1984xv,Bennett:1987vf,Allen:1990tv}. Global strings produced by the breaking of a global symmetry mainly radiate Goldstone bosons at the expense of gravitational radiation. 
Since we are interested only in gravitational wave (GW) signals, we focus on gauge strings and superstrings, and describe them as Nambu-Goto strings.

Both cosmic superstring and non-Abelian gauge string networks may have multiple string types. However, since gauge strings are often realized as topological defects of gauged $U(1)$ symmetry breaking in grand unified theories~\cite{Jeannerot:2003qv}, we focus on gauge string networks with a single string type. Throughout, we associate superstrings with multi-string networks and gauge strings with single-string networks.

A fundamental difference between gauge string and superstring networks is in their evolution.
As a cosmic network evolves, strings cross and reconnect by exchanging string segments (intercommute) with probability $P$ or pass through each other with probability $1-P$.
Loops formed in this process may then reconnect to the network or become isolated. The isolated loops oscillate under their tension $\mu$ and emit GWs with a signal strength proportional to their number density.
Because the intercommutation probability is unity for gauge strings and $P<1$ for superstrings \cite{Jackson:2004zg}, the superstring network gives a stronger GW signal.
Another complication in the evolution of a superstring network comes from the interaction of the different types of strings in the network. 
The intercommutation probabilities of F-strings, D-strings and their bound states, FD-strings, with themselves and with each other are generally not equal, and can differ by a couple of orders of magnitude. 

Recent data from pulsar timing arrays (PTAs) at nHz frequencies \cite{NANOGrav:2023gor,Antoniadis:2023ott,Reardon:2023gzh,Xu:2023wog} suggest that gauge string networks
do not fit the data because they yield a GW 
spectrum that is either too flat or too weak. The suppressed intercommutation probability for superstrings allows a larger amplitude and accommodates the observed blue-tilted GW spectrum. For related work in the context of PTA data see Refs.~\cite{NANOGrav:2023hvm,Ellis:2023tsl, Ellis:2020ena, Blasi:2020mfx,Blanco-Pillado:2021ygr}.
However, cosmic strings produce an observable GW spectrum over many orders of magnitude in frequency.
To conclusively infer a preference for a gauge string or superstring network, it will be necessary to consider the data from numerous planned interferometers including space-based laser interferometers (LISA \cite{Audley:2017drz}, Taiji \cite{Guo:2018npi}, TianQin \cite{Luo:2015ght}, BBO \cite{Corbin:2005ny}, DECIGO \cite{Seto:2001qf}, $\mu$Ares \cite{Sesana:2019vho}), atomic interferometers (MAGIS \cite{Graham:2017pmn}, AEDGE \cite{Bertoldi:2019tck}, AION \cite{Badurina:2019hst}), and ground-based interferometers (Einstein Telescope \cite{Sathyaprakash:2012jk} (ET), Cosmic Explorer \cite{Evans:2016mbw} (CE)), and Square Kilometre Array \cite{Janssen:2014dka} (SKA). 

The goal of our work is to perform a careful study of GW signals from cosmic superstring and gauge string networks. We study the impact of deviations from the scaling regime and the evolving number of relativistic degrees of freedom on the GW spectrum for gauge string networks in Section~\ref{sec2} (and generalize this to superstring networks in Section~\ref{sec3}).
 In Section~\ref{sec3}, we perform a detailed modeling of the superstring network, beyond the approximation of scaling the amplitude of the GW spectrum for gauge strings by $1/P$. We include multiple string types, their unequal intercommutation probabilities, the transition efficiencies between different string types, and the volume suppression because the strings evolve in a higher dimensional space. In Sections~\ref{sec2} and~\ref{sec3}, we provide analytical expressions for the GW spectrum for the two kinds of networks that apply for a wide frequency range.
In Section~\ref{sec4}, we analyze recent PTA data from NANOGrav~\cite{NANOGrav:2023gor}, EPTA/InPTA~\cite{Antoniadis:2023ott} and PPTA~\cite{Reardon:2023gzh}. We also consider how future data can discriminate between superstring and gauge string networks. We summarize in Section~\ref{sec5}.

\section{Cosmic gauge strings and GWs}
\label{sec2}

\subsection{Dynamics of long strings and loops}

 We begin with a brief review of the dynamics of Nambu-Goto strings. 
We only consider the network evolution from the radiation dominated era to today. The Hubble expansion rate $H\equiv \dot{a}/a$ can be written in terms of the matter and radiation density parameters and redshift $z$ as
\begin{eqnarray}
H(z) = H_0 \Big[1-\Omega_m-\Omega_r + (1+z)^3 \Omega_m + \mathcal{C}(z) (1+z)^4 \Omega_r\Big]^{\frac12}\,,
\end{eqnarray}
where
\begin{eqnarray} \label{eq:g_eff}
\mathcal{C}(z)=
 \frac{g_* (z)}{g_*(0)} \left(\frac{h_*(z)}{h_*(0)}\right)^{-\frac43}\,,
\end{eqnarray}
is a correction factor that accounts for the fact that $aT \propto h_*^{-\frac 13}(z)$ and is not constant.
Here, $g_*$ and $h_*$ are the energy density and entropy degrees of freedom, respectively. For the Standard Model, the redshift dependence of these parameters are provided in 
{\tt MicrOMEGAs\,5.2}~\cite{Belanger:2018ccd}. 
We find the following semi-analytical formula to be a good approximation to $\mathcal{C}(z)$:
\begin{eqnarray}
    \mathcal{C} (z) = 1 - 0.05\sigma(29.30 + 8.73x) - 0.12\sigma(38.50 + 10.08x) - 0.09\sigma(12.89 + 9.29x) \nonumber\\- 0.28\sigma(10.14 + 12.70x) - 0.042\sigma(-0.52 + 5.66x) - 0.03\sigma(-7.64 + 4.82x) \,,
\end{eqnarray}
where $x(z)=\ln[(1+z)\frac{T_0}{\rm GeV}]$ with $T_0$ the temperature of the Universe today, and $\sigma(y) = 1/(1+e^{-y})$. 
For $z \lesssim 10^8$, i.e., for temperatures less than tens of keV, $\mathcal{C}$ can be fixed at unity. 
For $z \gtrsim 2 \times 10^{14}$, corresponding to temperatures above the electroweak scale, $\mathcal{C} =0.388$. The value of $\mathcal{C}$ can fall at higher redshifts if new degrees of freedom contribute to the thermal bath at higher temperatures. This suppresses the GW spectrum above 10 Hz~\cite{Blanco-Pillado:2017oxo,Cui:2018rwi,Auclair:2019wcv}, and makes it easier to achieve consistency with the LIGO-Virgo-KAGRA (LVK) bound~\cite{LIGOScientific:2021nrg}.  However this suppression is not very significant unless $\mathcal{C}$ changes dramatically since, as we show in Section~\ref{eq:analytical_high_frequency}, the GW energy density at high frequencies depends linearly on $\mathcal{C}$. For example, in the MSSM, $g_*$ and $h_*$ are doubled at sufficiently high temperatures, which leads to the GW signal being suppressed by a factor of $1/2^{1/3} \sim 0.8$ at high frequencies. 
We do not consider such a suppression from new degrees of freedom.

We work with the velocity-dependent one-scale model~\cite{Martins:1995tg,Martins:1996jp} which describes network evolution in terms of a characteristic length scale $L$ of the network, called the correlation length, and the root-mean-square velocity $v$ of string segments. 
In the one-scale model, $L$ is the average radius of curvature of strings and the average inter-string distance, although these quantities are different in general.
On large scales, with one string segment of length $L$ in a volume $L^3$, the energy density of the network is $\rho = \mu/L^2$. The energy density evolves in the expanding Universe as 
\begin{eqnarray} \label{eq:drhodt}
\dot{\rho} &=& - \left[ 2 H (1+v^2)  + \frac{v^2}{\ell_f} + \frac{ \tilde{c}\, v}{L} \right] \rho  \,.
\end{eqnarray}
Here,  
$\ell_f$ is the friction length scale assumed to be infinity in the following. 
The last term accounts for energy loss into loops with $\tilde{c}$ the efficiency of chopping loops from the network, and found numerically to be $\tilde{c} \simeq 0.23$~\cite{Martins:2000cs}.
We do not consider the back reaction of loops on long strings. 
The density evolution depends on $v$, which satisfies
\begin{eqnarray} \label{eq:dvdt}
\dot{v} &=& (1- v^2) \left[ \frac{k(v)}{L} - 2 H v \right]\,,
\end{eqnarray}
where the momentum or curvature parameter~\cite{Martins:2000cs}
\begin{eqnarray}
k(v) = \frac{2\sqrt{2}}{\pi} (1-v^2) (1+2\sqrt{2} v^3)\, \frac{1-8 v^6}{1+8 v^6} \,,
\end{eqnarray}
accounts for acceleration due to the curvature of the strings. 
For the values of $v$ of interest, $v \in [0.55,0.75]$, 
$k(v) \simeq \frac{12}{\pi}(\frac{1}{\sqrt{2}}-v)$ is a good approximation. 

Since the network evolves to a linear scaling regime in which $L$ is constant relative to the horizon distance $d_H=2t$ ($d_H=3t$) in the radiation (matter) era,  it is convenient to normalize the correlation length as $\xi = L/t = \sqrt{\mu/\rho}/t$. The evolution equations~\eqref{eq:drhodt} and \eqref{eq:dvdt} become
\begin{eqnarray}
t\,\dot{\xi} &=&\beta (1+v^2) \xi - \xi + \frac{1}{2} \tilde{c} v \,, \nonumber\\
t\,\dot{v} &=& (1- v^2) \left[ \frac{k(v)}{\xi} - 2 \beta v \right]\,,
\end{eqnarray}
where $\beta = Ht$ is approximately constant: $\beta = \frac12$ in the radiation era if the variation of $g_*$ is neglected; $\beta = \frac23$  in the matter era. Numerical calculation shows a constant solution for $\xi$ and $v$. This is analytically confirmed by assuming $\dot{\xi} = \dot{v} = 0$, and leads to 
\begin{eqnarray} \label{eq:scaling}
\xi_{r} &=& 0.271\,, \quad v_{r} = 0.662\,, \quad \text{ radiation dominated era} \nonumber \\
\xi_{m} &=& 0.625\,, \quad v_{m} = 0.582\,, \quad \text{matter dominated era} \,
\end{eqnarray}

An increase in the loop energy density $\rho_{\circ}$ results from string collisions and from strings bending back on themselves, and is given by the last term in Eq.~\eqref{eq:drhodt}~\cite{Kibble:1984hp}. We include the Lorentz factor $\gamma_v = (1-v^2)^{-1/2}$ to account for energy loss due to redshifting of the velocity of the loops.
We also include a factor ${\cal F} \simeq 0.1$ to mimic the GW emission obtained from a full loop distribution via numerical simulations with only large loops~\cite{Blanco-Pillado:2013qja}; see also Ref.~\cite{Sanidas:2012ee}. Then, assuming that all the energy
lost by the network ends up in loops as in Eq.~\eqref{eq:drhodt},
\begin{eqnarray} \label{eq:loop_production}
\dot{\rho}_{\circ}= {\cal F} \frac{\tilde{c} v}{\gamma_v L} \rho = {\cal F} \frac{\tilde{c} v}{\gamma_v} \frac{\mu}{\xi^3 t^3}\,.
\end{eqnarray}
We also consider the evolution of the number density $n(l,t)$ of loops of length $l$. The energy density of loops with lengths in the interval $(l, l+dl)$ is $\mu l \cdot n(l,t)\, dl$, which on integration gives
\begin{eqnarray}
\rho_{\circ} = \int \mu l \cdot n(l,t)\, dl \,.
\end{eqnarray}
Assuming that all loops are produced with a length that is a  constant fraction $\alpha_L$ of $\xi$, we obtain the loop production function,
\begin{eqnarray} \label{eq:loop_chop}
{\cal P}(l,t)= {\cal F} \frac{\tilde{c} v}{\gamma_v \alpha_L \xi^4 t^5} \, \delta \left(\alpha_L \xi - \frac{l}{t} \right)\,,
\end{eqnarray}
where the delta function guarantees that $l/t = \alpha_L \xi$.
${\cal P}(l,t) dl dt$ is the increase in the number of loops with lengths in the interval $(l, l+dl)$ in the time period $(t, t+dt)$ per unit volume.
Accounting for the Hubble expansion, we integrate ${\cal P}(l,t)$ from an initial time $t_{\rm ini}$ to $t$ to find the loop number density distribution function at $t$:
\begin{eqnarray}
n(l,t) = \int_{t_{\rm ini}}^t  \, {\cal P}(l'(t'),t') \left[\frac{a(t')}{a(t)}\right]^3 dt'\,,
\end{eqnarray}
where the loop length under the integral is $l'(t') = l+ \Gamma G \mu (t - t')$ for $t>t'$. Here, $\Gamma \simeq 50 $ is the total GW emission power in units of $G\mu^2$~\cite{Blanco-Pillado:2017oxo}.  Doing the integral analytically gives~\cite{Sousa:2013aaa,Auclair:2022ylu}, 
\begin{eqnarray}
t^4n(l,t) = {\cal F} \frac{\tilde{c} v_\star}{\gamma_{v\star} \alpha_L \xi^4_\star} \, 
\frac{1}{  \alpha_L \xi_\star + \alpha_L \dot{\xi}_\star t_\star+\Gamma G \mu} \, 
\left[\frac{a(t_\star)}{a(t)}\right]^3 \,
\left[\frac{t}{t_\star} \right]^4 \,.
\end{eqnarray}
Here, $t_\star$ is the time when a loop is produced from the string network, and quantities with a $\star$ subscript are evaluated at $t=t_\star$. In particular, $\alpha_L \xi_\star= l_\star / t_\star$, where $l_\star$ is the length of the loop at formation; it is reasonable that the loop length at formation is proportional to the correlation length at formation. $t_\star$ is determined by
\begin{eqnarray}
\alpha_L \xi_\star t_\star + \Gamma G\mu t_\star = l + \Gamma G \mu t\,,
\end{eqnarray}
via the delta function in Eq.~\eqref{eq:loop_chop}. 
In the radiation era, $a(t_\star)/a(t)  = (t_\star/t)^{\frac12}$, so that
\begin{eqnarray} \label{eq:number_distribution}
t^4n(l,t) = {\cal F} \frac{\tilde{c}v_\star}{\gamma_{v\star} \alpha_L \xi^4_\star} \, 
\frac{1}{   \alpha_L \xi_\star + \alpha_L \dot{\xi}_\star t_\star + \Gamma G \mu} \, 
\left[ \frac{\alpha_L \xi_\star + \Gamma G \mu}{l/t + \Gamma G \mu}\right]^{\frac52}\,.
\end{eqnarray} 
Note that $t^4 n(l,t)$ is dimensionless and nonzero only if $t_{\rm ini} < t_\star < t$. 
Further simplification is possible if the scaling solution i.e., $\dot{\xi}(t) = 0$, holds:
\begin{eqnarray} \label{eq:number_density_star}
t^4 n(l,t) = {\cal F} \frac{\tilde{c}v_\star}{\gamma_{v\star} \alpha_L \xi^4_\star} \, 
\frac{(\alpha_L \xi_\star + \Gamma G \mu)^{\frac32}}{(l/t + \Gamma G \mu)^{\frac52}}\,.
\end{eqnarray}
If $\alpha_L \xi_\star \simeq 0.1 \gg \Gamma G \mu$,
then using the numerical values for $\xi$ and $v$ in the radiation era in Eq.~\eqref{eq:scaling}, gives 
\begin{eqnarray} \label{eq:number_density}
t^4 n(l,t) \simeq \frac{0.18}{(l/t + \Gamma G \mu)^{\frac52}} \,.
\end{eqnarray} 
This is consistent with large-scale
simulations of a Nambu-Goto string network 
which find that in the scaling regime $n(l,t)$ has a polynomial dependence on $l/t+\Gamma G\mu$, and that $\alpha_L \xi_\star \simeq 0.1$ corresponds to the peak of the loop distribution function~\cite{Blanco-Pillado:2013qja}. 

\begin{figure}[t!]
\centering
\includegraphics[width=.75\textwidth]{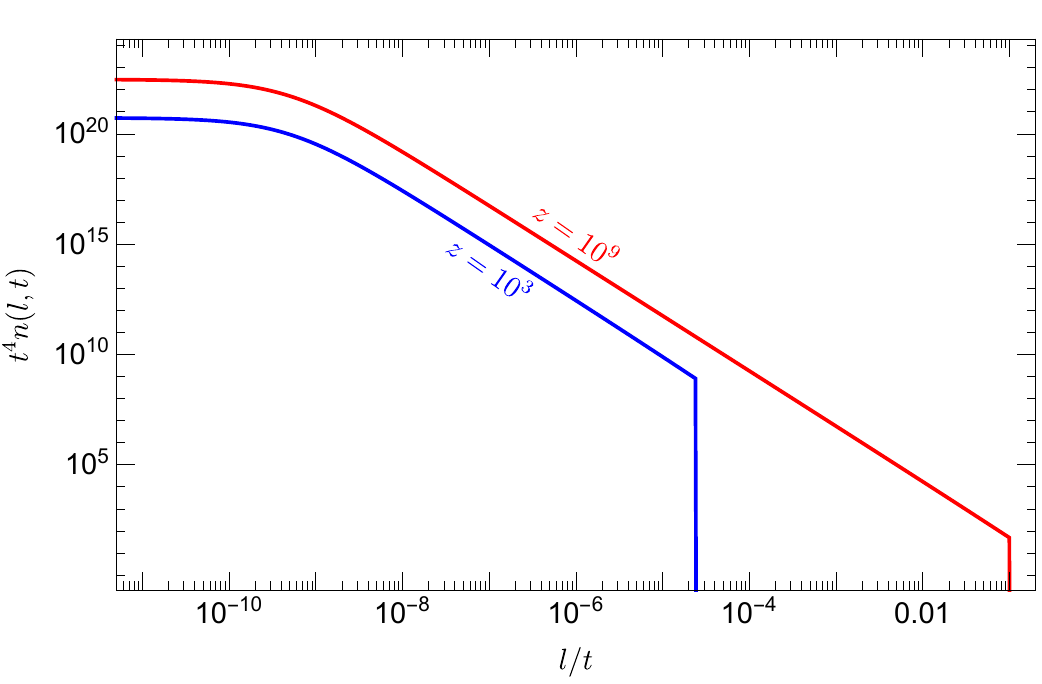}
\caption{Loop number density distribution $t^4 n(l,t)$ of gauge strings at redshifts $z= 10^9$ ($t=26~{\rm s}$), and $z=10^3$ ($t=4.2 \times 10^5~{\rm yr})$. The string tension is $G\mu = 10^{-11}$. }\label{fig:loop_density}
\end{figure}

Rather than fixing $\xi$ at its scaling solution, we consider the general formula in Eq.~\eqref{eq:number_distribution} with a fixed fraction $\alpha_L$ of loop length chopped from the long string. Taking $\alpha_L = 0.1/\xi_r$ corresponding to the peak of the standard scaling solution in the radiation era, we find
\begin{eqnarray} \label{eq:number_distribution_2}
t^4n(l,t) \simeq {\cal F} \frac{\tilde{c}v_\star}{\gamma_{v\star} \xi^3_\star} \, 
 \frac{0.32}{[1 + t_\star \, \dot{\xi}_\star/\xi_\star ][l/t + \Gamma G \mu]^{\frac52}}\,.
\end{eqnarray}  
Here, $t_\star$ can be approximately expressed in terms of $l/t$ as
\begin{eqnarray} 
t_\star \simeq \frac{t}{\alpha_L \xi_{\rm r}}( l/t + \Gamma G \mu) \,,
\end{eqnarray}  
where we have replaced $\xi_\star$ by its scaling value $\xi_{\rm r}$ as its variation with $t_\star$ is not significant in the radiation era. For illustration, in Fig.~\ref{fig:loop_density}, we show (with $t_{\rm ini}=0$) the loop number density as a function of the normalized loop length at redshift $z= 10^9$ ($t=26~{\rm s}$) in the radiation dominated era, and $z=10^3$, ($t=4.2 \times 10^5~{\rm yr} \simeq 1.4 \times 10^{13}~{\rm s}$) in the matter dominated era. We have compared $t^4 n(l,t)$ for the nonscaling solution and the scaling solution with varying $\mathcal{C}$ and found that  the relative error is less than $\sim 10\%$.

\subsection{GWs from string loops}
\label{sloop}

The spectrum of the gravitational wave background in terms of the frequency $f$ in the current epoch $t_0$ is
\begin{eqnarray}
\Omega_{\rm GW} (f) = \frac{1}{\rho_c} \frac{d\rho_{\rm GW}(t_0, f)}{d\ln f}\,,
\end{eqnarray}
where $\rho_{\rm GW}(t_0,f)$ is the energy density in GWs
and $\rho_c = 3 H_0^2 /(8\pi G)$ is the critical density. 
The relic background signal today from GWs of frequency $f'$ emitted from a gauge string loop at time $t$ must account for the redshift of the frequency and the $(1+z)^{-4}$ dilution of the density due to the Universe's expansion. 
Approximating the formation time of the loop to be $t=0$, the energy density in GWs today is given by
\begin{eqnarray} \label{eq:rho_GW}
\frac{d \rho_{\rm GW}(t_0, f)}{df}
= \int_0^{t_0} \frac{dt}{(1+z(t))^4} P_{\rm GW} (t,f') \frac{\partial f'}{\partial f}\,,
\end{eqnarray}
where
\begin{eqnarray}
    P_{\rm GW} (t,f') = G\mu^2 \sum_{k=1}^{\infty} \frac{l}{f'} n(l,t) P_k\,
\end{eqnarray} 
is the GW power at frequency $f'=(1+z)f$ from loops radiating at time $t$. 
The emitted frequency $f'=2k/l$ corresponds to oscillations of a loop of length $l$ in harmonic modes $k =1,2,3,\dots$, and 
$P_k$ is the normalized GW power (in units of $G\mu^2$) emitted by a loop in harmonic mode $k$. The GW energy density can be written as a summation of the harmonic series, 
\begin{eqnarray} \label{eq:rho_GW_2}
\frac{d \rho_{\rm GW}(t_0, f)}{df} = G\mu^2 \sum_{k} C_k(f) P_k\,,
\end{eqnarray}
where 
\begin{eqnarray} \label{eq:C_k}
C_k(f) = \frac{2k}{f^2} \int_0^{\infty} \frac{dz}{H(z)(1+z)^6} n\bigg(\frac{2k}{(1+z)f}, t(z)\bigg)
\end{eqnarray}
is the time-integrated weight function of loops in mode $k$ that emit GWs detected with frequency $f$. We only include the GW contribution from cusps (pieces of loops where the string bends back on itself) since it dominates over kinks (sharp structures at points where strings reconnect after colliding). Then $P_k$ is given by 
\begin{eqnarray}
P_k \simeq \frac{\Gamma k^{-\frac43}}{\zeta(\frac43)}\,,
\end{eqnarray}
where $\zeta(q) = \sum_{k=1}^\infty k^{-q}$ is the Riemann zeta function. 
Strictly, without including backreaction, the summation over $k$ should extend to infinity. We sum modes up to $k=10^4$ to keep the error under control.

\begin{figure}[t]
\centering
\includegraphics[width=.8\textwidth]{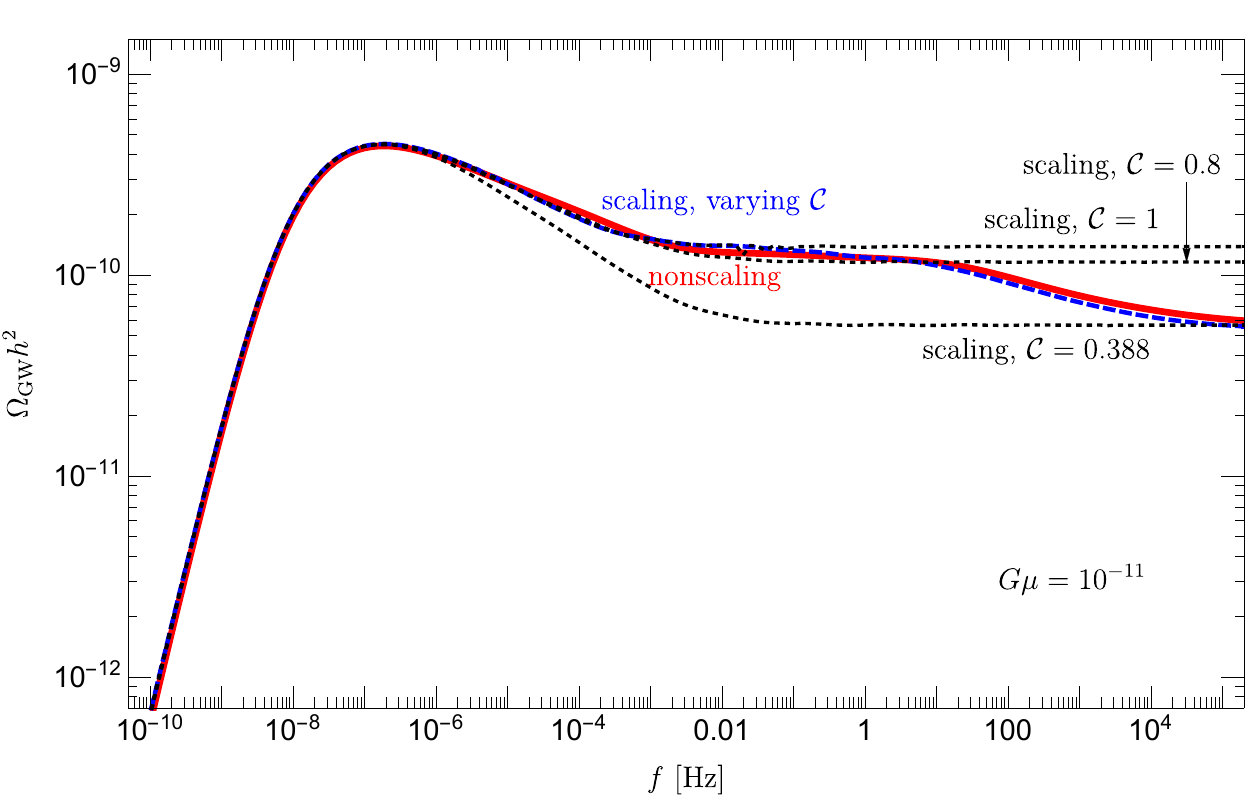}
\caption{\label{fig:GW_string}
GW spectrum calculated using three procedures for evaluating the loop number density.
The spectrum labeled ``nonscaling" does not assume that the cosmic string network is in the scaling regime, and is most precise; see Eq.~\eqref{eq:number_distribution_2}. 
The other four spectra are for the case of a scaling network; see Eq.~\eqref{eq:number_density_star}. 
The label ``varying $\mathcal{C}$'' means that the correlation length $\xi_\star$ varies with time because $\mathcal{C}$ evolves with redshift according to Eq.~\eqref{eq:g_eff}. The spectra for two oversimplified scaling cases in which 
$\mathcal{C}$ is fixed to constant values in the radiation era are also shown. The spectrum labeled ``$\mathcal{C}=1$'' assumes $g_* = 3.37$, the number of relativistic degrees of freedom after BBN, and ``$\mathcal{C}=0.388$'' assumes $g_*= 106.75$, i.e., all Standard Model particles contribute to the thermal plasma. We find 
empirically that ``$\mathcal{C}=0.8$'' matches the nonscaling solution between $10^{-3}$~Hz and 10~Hz.} 
\end{figure}

The GW spectrum $\Omega_{\rm GW}(f)h^2$ for $G\mu = 10^{-11}$ for
three different treatments of the loop number density distribution is shown in Fig.~\ref{fig:GW_string}:
\begin{enumerate}
    \item The solid curve is the GW spectrum for the nonscaling solution in Eq.~\eqref{eq:number_distribution_2} which accounts for the variation of the correlation length and $\mathcal{C}$ as the Universe expands. Since $t^4 n(l,t)$ varies with $\xi$, its value at loop production $t_\star^4 n(l_\star,t_\star)$, can differ from its value during GW emission. 
    \item The dashed curve is the GW spectrum for the scaling solution in Eq.~\eqref{eq:number_density_star} with $\xi$ fixed at $\xi_\star$ (or equivalently $l_\star/t_\star = \alpha_L \xi_\star$). In this case $t^4 n(l,t)$ is the same at loop production and during GW emission. However, as the Universe expands, $\mathcal{C}$ decreases and consequently $\xi_\star$ and $t^4 n(l, t)$ vary with time on a timescale much greater than that of GW emission.
    \item The dotted curve is the GW spectrum for the oversimplified scaling solution in  Eq.~\eqref{eq:number_density} with $\xi = \xi_{\rm r}$ and $\alpha_L \xi_\star=0.1$. This is achieved by fixing $\mathcal{C} = 0.388$, which means we do not vary the number of relativistic species in the thermal universe. We also include the spectrum for 
    $\mathcal{C} = 1$ corresponding to the number of relativistic species after the end of Big Bang nucleosynthesis (BBN), and the spectrum for an empirically obtained value $\mathcal{C} = 0.8$ that matches the nonscaling solution between $10^{-3}$~Hz and 10~Hz.
    
\end{enumerate}
Of these solutions, the first gives the most precise description of the GW spectrum but is computationally intensive; the last is easiest to compute but provides the worst description of the GW spectrum. We now discuss which procedure is most suitable for describing data from current and future PTAs and interferometers. Note the following from Fig.~\ref{fig:GW_string}.
\begin{enumerate}
\item
The 1 - 100 nHz band is preferred by PTAs. In this band, all three procedures agree very well with each other. The oversimplified scaling solution is adequate to describe the GW spectrum.

\item 
Space-based laser interferometers can measure GWs in the mHz - Hz range. While the oversimplified scaling solution deviates substantially from the other two solutions, the scaling solution with varying $\mathcal{C}$ agrees quite well with the nonscaling solution. So, if there is not a strong requirement on precision, the scaling solution with varying $\mathcal{C}$ can serve as an alternative to the nonscaling solution.

\item 
In the Hz - kHz range preferred by ground-based experiments, the nonscaling solution and the scaling solution with varying $\mathcal{C}$ match very well. The oversimplified scaling solution does a poor job. 

\end{enumerate}

The reason why the oversimplified scaling solution deviates from the other two solutions in a broad range of frequencies is as follows.
In both the scaling solution with varying $\mathcal{C}$ and the nonscaling solution, the varying $\mathcal{C}$ leads to a varying $\xi$, and consequently a varying $t^4 n(l,t)$. If $\xi$ varies slightly from the time of loop production to the time of GW emission, one expects the loop number densities calculated in these two ways to have smaller deviations than that calculated via the oversimplified scaling solution with the correlation length fixed at $\xi_{\rm r}$.

\subsection{Analytical expressions for the GW spectrum}
\label{sec:GW_formulism}

\subsubsection{PTA band}
We derive an analytical expression for the GW spectrum in the 1 - 100~nHz band relevant to PTAs. As demonstrated in the last subsection, the oversimplified scaling solution gives a precise description of the GW spectrum in this frequency range. We make two more assumptions: 
1) most loops are formed in the radiation era, and
2) GWs are mainly radiated from loops in the matter and dark energy dominated eras, during which the redshift-time relation (neglecting $\Omega_{r}$) is
\begin{eqnarray}
\sqrt{1 + (1+z)^3 \frac{\Omega_m}{1-\Omega_m}} = \frac{\exp (3 t H_0 \sqrt{1-\Omega_m} ) + 1}{\exp (3 t H_0 \sqrt{1-\Omega_m} ) - 1} \,.
\end{eqnarray}
The number density of loops produced in the radiation era and that survive until the latter eras is 
\begin{eqnarray}
    n_{\rm r,{\rm late}}(l,t) &=& \Big(\frac{1+z}{1+z_{\rm eq}} \Big)^3 n_{\rm r}(l_{\rm eq}, t_{\rm eq}) \simeq \Big(\frac{1+z}{1+z_{\rm eq}} \Big)^3 t_{\rm eq}^{-\frac32} \frac{0.18}{(l_{\rm eq} + \Gamma G \mu t_{\rm eq})^{\frac52}} \nonumber\\
    &\simeq& \Big(\frac{1+z}{1+z_{\rm eq}} \Big)^3 t_{\rm eq}^{-\frac32} \frac{0.18}{(l + \Gamma G \mu t)^{\frac52} } \,, 
\end{eqnarray}
where $t_{\rm eq}$ and $z_{\rm eq}$ define the epoch of matter-radiation equality, $l_{\rm eq} = l + \Gamma G\mu (t- t_{\rm eq})$, and we have used the scaling solution for $n(l,t)$ in Eq.~\eqref{eq:number_density}, which in the epoch of matter-radiation equality requires $l_{\rm eq}/t_{\rm eq} <0.1$. 

We first consider the case in which the evaporation rate $\Gamma G\mu \ll l/t$. Then, dropping the $G\mu$ contribution to the number density, Eq.~\eqref{eq:C_k} gives
\begin{eqnarray}
C_k^{(0)}(f) &=& \frac{1}{f} \Big(\frac{f}{2k}\Big)^\frac32  \frac{0.18}{(1+z_{\rm eq})^3 t_{\rm eq}^{\frac32} H_0}  B^{(0)}
\end{eqnarray}
with 
\begin{eqnarray}
B^{(0)}  =
\int_0^{\infty} \frac{dz}{\sqrt{(1+z)(1-\Omega_m) + (1+z)^4 \Omega_m}}
= \frac{1}{\sqrt{\Omega_m}}\,
_2 F_1 (\frac13, \frac12, \frac 43 ; \frac{\Omega_m-1}{\Omega_m})\,.
\end{eqnarray}
Here, ${}_2 F_1 (a,b,c;d)$ is the Hypergeometric function. 
By taking the Planck measurement, $\Omega_m = 0.3081 \pm 0.0065$~\cite{Planck:2018vyg}, we obtain $B^{(0)} = 1.519 \mp 0.011$. As $C_k \propto f^{1/2}$, the GW spectrum follows a power-law behaviour $\Omega_{\rm GW} \propto f \cdot f^{1/2} = f^{3/2}$. 
The GW spectrum for the best-fit $\Lambda$CDM cosmology is numerically given by
\begin{eqnarray} \label{eq:Omega_hsq_PTA_power_law}
\Omega_{\rm GW}(f) h^2 |_{\text{power-law,PTA}} = 
4.2 \times 10^{-9} \, \Big(\frac{f}{f_{\text{yr}}}\Big)^{3/2} \frac{\Gamma}{50} \, \Big(\frac{\text{G$\mu $}}{10^{-11}}\Big)^2 \sum_{k} \frac{k^{-\frac{17}{6}}}{\zeta(\frac{17}{6})}\,,
\end{eqnarray}
where $f_{\rm yr}= (1\ \rm yr)^{-1} \simeq 32$~nHz.
Note that the sum over $k$-dependent terms, $\sum_{k=0}^{\infty} \frac{k^{-q}}{\zeta(q)} = 1$, and is kept only to emphasize that the sum converges. 

Now we include the contribution of $G\mu$ to the number density. $C_k$ in the nHz band is given by
\begin{eqnarray}
C_k (f) &=& \frac{1}{f} \Big(\frac{f}{2k}\Big)^\frac32  \frac{0.18}{(1+z_{\rm eq})^3 t_{\rm eq}^{\frac32} H_0}  B_k \,,
\end{eqnarray}
where
\begin{eqnarray}
B_k = \int_0^{\infty} \frac{dz}{\sqrt{(1+z)(1-\Omega_m) + (1+z)^4 \Omega_m}} \Big[ 1 + u_k  (1+z) \Big]^{-\frac52}
\label{Bk}
\end{eqnarray}
with
\begin{eqnarray}
u_k = \frac{f}{2k}\frac{\Gamma G \mu}{3H_0 \sqrt{1-\Omega_m}} = \frac{2.89}{2k} \frac{f}{f_{\rm yr}} \, \frac{\Gamma}{50} \, \frac{G\mu}{10^{-11}}\,.
\end{eqnarray} 
This integral is not analytically solvable. However, we can use the semi-analytical formula,
\begin{eqnarray}
B_k \approx B^{(0)} \frac{1}{(1+ a u_k)^{b}}\,,
\label{Bkana}
\end{eqnarray} 
to fit the numerical result for $B_k$. Here we take $a = 2.075$ and $b = 1.945$. 
Fixing $\Omega_m=0.3081$, we find the relative error to be $B_k{\rm{(Eq.~\ref{Bk})}} / B_k{\rm{(Eq.~\ref{Bkana})}} -1 \lesssim 5\%$ for $u_k$ in a wide range, $0 < u_k < 10^3$. Relaxing $\Omega_m$ in the $1\sigma$ range, the relative error is within $10\%$. The requirement $u_k < 10^3$ translates to a requirement on the physical parameters:
\begin{eqnarray}
 \frac{f}{f_{\rm yr}} \, \frac{\Gamma}{50} \frac{G\mu}{10^{-11}} < 700\,.
\end{eqnarray}
Finally, we obtain 
\begin{eqnarray} \label{eq:Omega_hsq_PTA_approx}
\Omega_{\rm GW}(f) h^2|_{\rm PTA} &=& \frac{8 \pi  h^2}{3 H_0^3}(G\mu)^2 \frac{0.18}{(z_{\rm eq}+1)^3 t_{\text{eq}}^{\frac 32}}\sum _{k=1}^{\infty } \left(\frac{f}{2 k}\right)^{3/2}B_k P_k  \nonumber\\
&\approx& \Omega_{\rm GW}(f) h^2 |_{\text{power-law,PTA}} \times
\sum _{k=1}^{\infty } \frac{1}{(1+ 2.075 u_k){}^{1.945}} \, \frac{k^{-\frac{17}{6}}}{\zeta(\frac{17}{6})}\,.
\end{eqnarray}

\begin{figure}[t!]
\centering
\includegraphics[width=.45\textwidth]{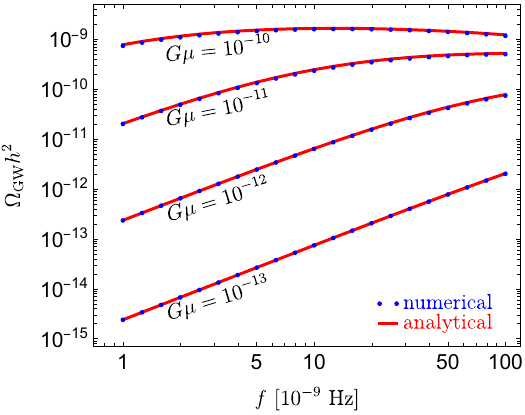}
\includegraphics[width=.45\textwidth]{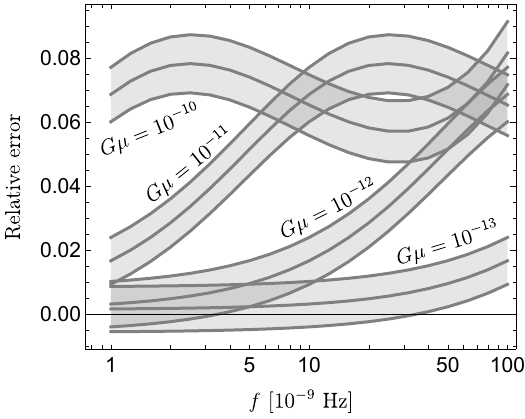}
\caption{Comparison between the numerical and semi-analytical calculations of the GW spectrum in the nHz band. Left panel: GW spectra calculated 
numerically using Eq.~\eqref{eq:rho_GW_2} 
 (blue dots) and analytically using Eq.~\eqref{eq:Omega_hsq_PTA_approx} (red curves) for $\Omega_m = 0.3081$. Right panel: 
 Relative error in Eq.~\eqref{eq:Omega_hsq_PTA_approx}
 for $\Omega_m = 0.3081 \pm 0.0065$. } \label{fig:Omega_hsq_approx}
\end{figure}

In the nHz band, GW power in the $k$ mode is suppressed by $k^{-\frac{17}{6}}$, so the contribution from higher harmonic modes is highly suppressed. Thus, in numerical calculations it suffices to sum up to $k=50$.
As can be seen from Fig.~\ref{fig:Omega_hsq_approx}, 
this formula is a very good approximation to Eq.~\eqref{eq:rho_GW_2}.

We compare our analytical expression with the widely used model-independent power-law approximation, which applies in a very narrow frequency band. Given a band around a reference frequency $f_{\rm ref}$, the spectrum can be approximated by~\cite{Fu:2022lrn}
\begin{eqnarray} \label{eq:Omegahsq}
 \Omega_{\rm GW}(f) h^2 |_{\text{power-law}} \approx 2.02\cdot 10^{-10} \left(\frac{f_{\rm ref}}{f_{\rm yr}}\right)^{5-\gamma}\times \left(\frac{A}{10^{-15}}\right)^2 \left( \frac{f}{f_{\rm ref}} \right)^{5-\gamma}\,,
\end{eqnarray}
 where $\gamma = 3-2\alpha$. Here, $\alpha$ is the spectral index of the characteristic strain $h_c(f) = A \times (f/f_{\rm ref})^\alpha$. 
By fixing $f_{\rm ref}= f_{\rm yr}$ and comparing with Eq.~\eqref{eq:Omega_hsq_PTA_approx}, we obtain
\begin{eqnarray}
    A_{\rm PTA} &\simeq& 4.6 \cdot 10^{-15} \times 
    \frac{\Gamma}{50} \, \frac{G\mu}{10^{-11}} 
    \Big[ \sum_k \frac{k^{-\frac{17}{6}}}{(1+a u_k)^b \, \zeta (\frac{17}{6})} \Big]^{\frac12}\Big|_{f=f_{\rm yr}}\,, \nonumber\\
    \gamma_{\rm PTA} &\simeq& \frac72 + \frac{ab}{\sum_{k'} {k'}^{-\frac{17}{6}} B_{k'}} \sum_k \frac{k^{-\frac{17}{6}} u_k B_k }{(1+a u_k)} \Big|_{f=f_{\rm yr}}\,,
\end{eqnarray}
which implies $\gamma_{\rm PTA} \geqslant 7 /2$. In the case of small $G\mu$, $u_k$ can be neglected, and we recover the power-law spectrum in Eq.~\eqref{eq:Omega_hsq_PTA_power_law} with 
\begin{eqnarray}
    \gamma_{\rm PTA} = \frac72\,, \quad
    A_{\rm PTA} = 4.6 \cdot 10^{-15} \times 
    \frac{\Gamma}{50} \, \frac{G\mu}{10^{-11}}  \,.
\end{eqnarray}

\subsubsection{High frequency band} \label{eq:analytical_high_frequency}

We derive analytical formulas for the GW spectrum in the high frequency range,  $10^{-3} - 10^{3}$~Hz, which can be detected by space- and ground-based laser inteferometers. 

It is known that signals at these frequencies arise from GWs emitted in the deep radiation era. Thus, the upper limit of integration in Eq.~\eqref{eq:rho_GW} is $\tilde{t}$, a time much before matter-radiation equality $t_{\rm eq}$. In this case, 
\begin{eqnarray}
    H(\tilde{z}) = H_0 \sqrt{\mathcal{C}\Omega_R} (1+\tilde{z})^2 \,, \quad
    \tilde{t} = \frac{1}{2 H_0 \sqrt{\mathcal{C}\Omega_R} (1+\tilde{z})^2}\,.
\end{eqnarray} 
Here, we treat $\mathcal{C}$ as a constant although it varies from 0.388 to 1. The GW energy density today can be written in the form,
\begin{eqnarray}
    \rho_{\rm GW}(t_0, f) = \frac{\rho_{\rm GW}(\tilde{t}, \tilde{f})}{(1+\tilde{z})^4}\,,
\end{eqnarray}
where $\tilde{f} = (1+\tilde{z})f$ is the frequency emitted at $\tilde{t}$ corresponding to redshift $\tilde{z}$.

Following Section~\ref{sloop}, we write the GW spectrum at time $\tilde{t}$ as 
\begin{eqnarray}
    \frac{d \rho_{\rm GW}(\tilde t, \tilde f)}{df} = G \mu^2 \sum_k  P_k \frac{2k}{\tilde{f}^2} \times \int_0^{\tilde{t}} \frac{dt}{(1+\tilde{z})^5} n(l,t)\,,
\end{eqnarray}
where $l$ is related to $\tilde{f}$ via $l = \frac{2k}{(1+z(t))\tilde{f}}$. 
Using the scaling solution in Eq.~\eqref{eq:number_density}, the integral can be solved explicitly:
\begin{eqnarray}
    \int_0^{\tilde{t}} \frac{dt}{(1+\tilde{z})^5} n(l,t) = \frac{1}{(1+\tilde{z})^5}\frac{0.24}{\tilde{t}^{1/2}} \Big( \frac{\tilde{f}}{2k} \Big)^{\frac52} (1+\tilde{b})^{-\frac32}\,,
\end{eqnarray}
where $\tilde{b} = \tilde{t} \, \Gamma G\mu \, \tilde{f} / (2k)$. In the high frequency band, $\tilde{b} \gg 1$, we find
\begin{eqnarray}
    \frac{d \rho_{\rm GW}(\tilde t, \tilde f)}{df} = \frac{0.24}{G \tilde{f} \tilde{t}^2} \, \sqrt{\frac{G\mu}{\Gamma}} \, \sum_k \, \frac{k^{-\frac43}}{\zeta(\frac43)}\,.
\end{eqnarray}
Since $\tilde{f} \rho_{\rm GW}(\tilde{t}, \tilde{f})$ is independent of $\tilde{f}$, the GW spectrum today is also frequency-independent:
\begin{eqnarray} \label{eq:Omega_hsq_laser}
    \Omega_{\rm GW}(f) h^2 |_{\rm laser} &=& 2.56 \pi  \, \mathcal{C} \Omega_r h^2\, \sqrt{\frac{G\mu}{\Gamma}} \, \sum_k \frac{k^{-\frac43}}{\zeta(\frac43)} \nonumber\\
    &\simeq& 4.78 \cdot 10^{-5} \times \mathcal{C} \sqrt{G\mu}\,.
    \label{eq:laser}
\end{eqnarray}
In the first of these equations, we have again kept the sum over $k$ modes although it is unity. By taking $\mathcal{C} =1$, we recover the result of Ref.~\cite{Blanco-Pillado:2017oxo}. With ${\cal C} = 1$ and ${\cal C} = 0.388$, this formula reproduces the plateau of the dotted curves in Fig.~\ref{fig:GW_string} perfectly.  However, these spectra deviate from the GW spectrum for the nonscaling solution. The GW spectrum obtained with ${\cal C}= 0.8$ matches the nonscaling solution better in a wide frequency range, $1~{\rm mHz}\lesssim f \lesssim 10~{\rm Hz}$. This empirically obtained value of ${\cal C}$ is almost independent of $G\mu$.

The power-law approximation in Eq.~\eqref{eq:Omegahsq}
is valid with $\gamma=5$. Then, the spectrum can be approximated by
\begin{eqnarray} \label{eq:Omegahsq_laser}
\Omega_{\rm GW}(f) h^2 |_{\text{laser}} \approx 2.02\cdot 10^{-10} \left(\frac{A_{\rm laser}}{10^{-15}}\right)^2 
\end{eqnarray}
with
\begin{eqnarray}
    A_{\rm laser} = 8.65 \cdot 10^{-16}  \times \sqrt{{\cal C}} \Big( \frac{G\mu}{10^{-11}} \Big)^{1/4} \,.
\end{eqnarray}
Note that in the high-frequency band, $A_{\rm laser} \propto (G\mu)^{1/4}$, while in the low-frequency band $A_{\rm PTA} \propto G\mu$.

\subsubsection{Convergence at low and high frequencies}

From Eqs.~\eqref{eq:Omega_hsq_PTA_approx} and \eqref{eq:laser}, the dependence of the GW spectrum on $k$ at low and high frequencies is
\begin{eqnarray}
    \Omega_{\rm GW}(f)h^2|_{\rm PTA} &\propto& \sum_{k} k^{-\frac{17}{6}} \,, \nonumber\\
    \Omega_{\rm GW}(f)h^2|_{\rm laser} &\propto& \sum_{k} k^{-\frac{4}{3}}\,.
\end{eqnarray}
In the nHz range measured by PTA, the sum converges very fast. Summing up to $k =50$ keeps the relative error smaller than 0.04\%. 
Note that the requirement, $l_{\rm eq}/t_{\rm eq} <0.1$, is equivalent to $(1+z) > 2k/(0.1 f t_{\rm eq})$ in the small $\Gamma G \mu$ limit. For the lower end of the PTA frequency range, $f = 1$~nHz, the condition becomes $1+z > k/82$. Thus, summing up to $k=50$ allows for loops radiating at all possible redshifts. 
In the mHz - kHz range targeted by laser interferometers, the contribution of each $k$ mode to the GW signal is proportional to $k^{-4/3}$, so summing up to $k =50$ leads to a relative error larger than $20\%$. In our numerical calculations we sum up to $k = 10^4$, which gives a relative error smaller than $4\%$.


\section{Cosmic superstrings and GWs}
\label{sec3}

\subsection{Dynamics of long superstrings and loops}

We generalize our discussion to the superstring case which has different types of strings evolving in the Universe. 
In particular, we consider three types of strings with unequal tensions~\cite{Pourtsidou:2010gu,Avgoustidis:2007aa,Copeland:2009ga,Avgoustidis:2009ke,Sousa:2016ggw}:
\begin{eqnarray}
\text{ string~1~\hspace{1mm} (F-string)}:&& \mu_1 = \mu_F\,,\nonumber\\
\text{ string~2~\hspace{1mm} (D-string)}:&& \mu_2 = \mu_F /g_s\,,\nonumber\\
\text{ string~3~ (FD-string)}:&& \mu_3 = \mu_F \sqrt{1+1/g_s^2}\,,
\end{eqnarray} 
They correspond to the so-called $(p,q)$-cosmic strings with tensions given by $\mu(p,q) = \mu_F \sqrt{p^2+q^2/g_s^2}$, where $p$ is the number of quanta of F charge of the lightest fundamental (F-) string, $q$ is the number of quanta of D charge carried by D branes, $\mu_F$ is the tension of the F-string, and $g_s$ is the string coupling constant. $p$ and $q$ must be coprime numbers. The F-string, D-string and FD-string have $(p,q)=(1,0)$, $(0,1)$ and $(1,1)$, respectively. We refer to them as string 1, 2, and 3, respectively, with network energy density $\rho_i$, correlation length $L_i$ and mean velocity $v_i$ for $i=1,2,3$. Note that for very small values of $g_s$, the density of the F-strings becomes so high that the one-scale approximation breaks down. 

When two different types of strings collide, they may zip together to produce trilinear Y-junctions linked by a {\it zipper} made of the third type of string.
The evolution equation of $\rho_i$ includes the contribution of the newly produced type-$i$ segment from other strings and removes the contribution of the type-$i$ segment that produced other string types: 
\begin{eqnarray} \label{eq:drhodt_v2}
\dot{\rho_i} &=& - \left[ 2 H (1+v_i^2)  + \frac{ \tilde{c}_i v_i}{L} \right] \rho_i + \sum_{j,k}\dot{\rho}_{j,k \to i} - \sum_{j,k} \dot{\rho}_{i,j \to k} \,.
\end{eqnarray}
Here, $\dot{\rho}_{j,k \to i}$ is the production rate of a type-$i$ zipper from the collision of types-$j$ and -$k$ strings, and  $\dot{\rho}_{i,j \to k}$ is the annihilation rate of type-$i$ due to collisions with type-$j$ producing a zipper of type-$k$. The interaction must satisfy the selection rule, $p_i=(p_j+p_k,q_j+q_k)$ or $p_i=(|p_j-p_k|,|q_j-q_k|)$. 
As in the case of gauge strings, interactions among strings of the same type produce loops. $\alpha_L$ is assumed to be the same for all string types.
Energy conservation including the kinetic energies of the colliding strings requires that $\rho_i$ be corrected to \cite{Avgoustidis:2007aa}
\begin{eqnarray}
\rho_i =\frac{\mu_i}{L_i^2}\gamma_{v_i}\,,
\end{eqnarray}
where $\gamma_{v_i}=(1-v_i^2)^{-\frac12}$.
Changes in the string energy density contribute to the acceleration of the strings. 
The evolution of $\xi_i \equiv L_i/t$ and $v_i$ are governed by \cite{Avgoustidis:2007aa,Avgoustidis:2009ke}
\begin{eqnarray}
t\, \dot{\xi_i} &=& \beta (1+v_i^2) \xi_i - \xi_i + \frac{1}{2} \tilde{c}_i v_i + \frac{1}{2} \Big( \sum_{j,k} d_{ij}^k - \sum_{j<k}d_{jk}^i \Big) \xi_i^3 \,, \\\nonumber
t\, \dot{v}_i &=& (1- v_i^2) \left[ \frac{k(v_i)}{\xi_i} - 2\beta v_i + B \sum_{j,k} d_{jk}^i \frac{\mu_j+\mu_k-\mu_i}{\mu_i} \frac{\xi_i^2}{v_i} \right]\,.
\end{eqnarray}
where 
\begin{eqnarray}
d_{jk}^i = \frac{\tilde{d}_{jk}^i \sqrt{v_j^2 + v_k^2}}{\xi_j \xi_k (\xi_j +\xi_k)} \,.
\end{eqnarray}
The coefficient $B \in [0,1]$ is the fraction of energy  radiated away or transferred to the network as kinetic energy during Y-junction formation (since $\mu_i \neq \mu_j+\mu_k$ in general). We set $B=0$ (corresponding to all the energy being radiated away) since changing the value of $B$ does not significantly affect the dynamics of the network~\cite{Avgoustidis:2009ke}.
$\tilde{c}_i$ is the loop chopping efficiency for self-interactions of type-$i$ strings, and is defined as 
\begin{eqnarray}
    \tilde{c}_i = \tilde{c}\,P_{ii}^{\frac13}\,,
\end{eqnarray} 
with $\tilde{c}$ the chopping efficiency for gauge strings. $P_{ij}$ is the intercommutation probability for string $i$ and string $j$~\cite{Sousa:2016ggw}, and is generally not unity. It can be parameterized by the product of a volume-independent quantum interaction piece and a volume suppression factor $w\in (0,1]$ to account for the fact that the strings evolve in a higher-dimensional space. 
Geometrically, $w\simeq1$ corresponds to compactification at the string scale. Some key features of the intercommutation probabilities are~\cite{Jackson:2004zg,Hanany:2005bc}
\begin{enumerate}
    \item For F-F string interactions, $P_{11}$ is ${\cal{O}}(g_s^2)$ and takes values in the range $(10^{-3},1)$.
    \item For D-D or FD-FD string interactions, $P_{22}$ and $P_{33}$  $\in (0.1,1)$. 
    \item For F-D or F-FD string interactions, $P_{12}$ and $P_{13}$ are ${\cal{O}}(g_s)$, and take values in the range $(10^{-2},1)$.
\end{enumerate}
The cross-interaction efficiency $\tilde{d}_{jk}^i=\tilde{d}_{kj}^i$ is defined by 
\begin{eqnarray}
    \tilde{d}_{jk}^i = P_{jk}^{\frac13} S_{jk}^i \,,
\end{eqnarray}
where $S_{jk}^i$ is the conditional probability that a collision of type-$j$  and -$k$ strings produces a type-$i$ zipper~\cite{Pourtsidou:2010gu} given that an interaction has taken place.
The $g_s$-dependent chopping and transition efficiency factors $\tilde{c}_i$ and $\tilde{d}_{ij}^k$ obtained in Ref.~\cite{Pourtsidou:2010gu}  are listed in Table~\ref{Tab:super_string_chopping}. 

\begin{figure}[t!]
\centering
\includegraphics[width=.45\textwidth]{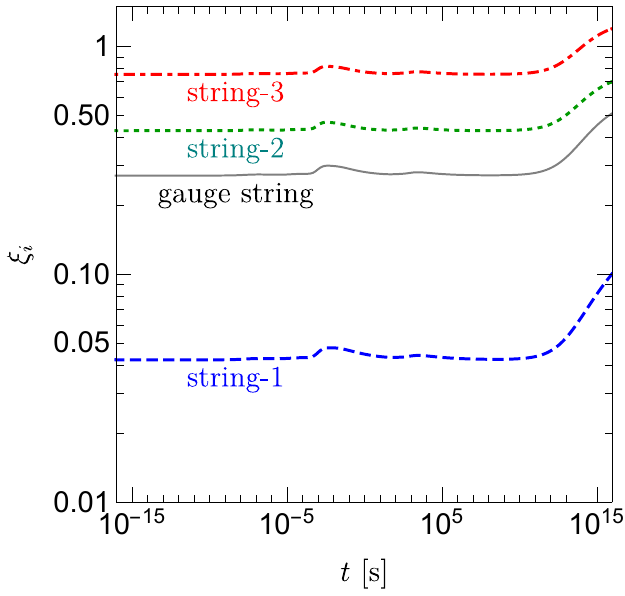}
\includegraphics[width=.45\textwidth]{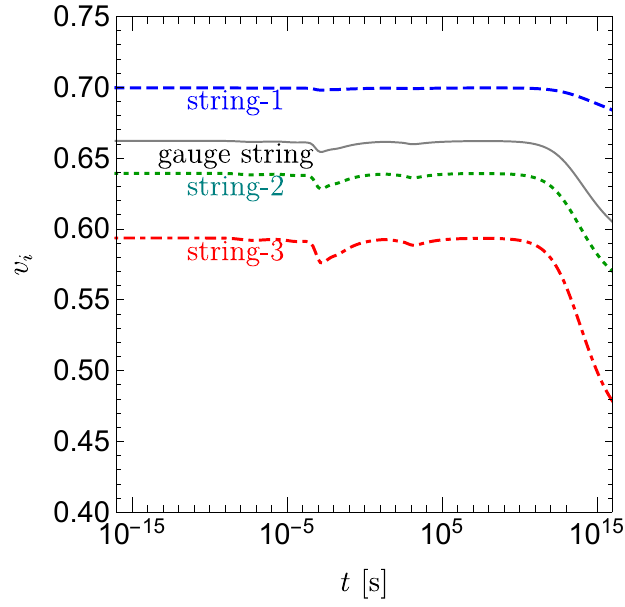}
\caption{Time evolution of normalized correlation lengths $\xi_i = L_i/t$ (left panel) and mean velocities $v_i$ (right panel) of superstrings for $i=1,2,3$. We take $g_s=0.1$, $w=1$ and $(\tilde{c}_1,  \tilde{c}_2,  \tilde{c}_3)$ and
$(\tilde{d}_{12}^3,\tilde{d}_{13}^2,\tilde{d}_{23}^1)$ from Table~\ref{Tab:super_string_chopping}. For comparison, the evolution of gauge strings is shown in gray.}\label{fig:scaling}
\end{figure}

In superstring networks, correlation lengths for different types of strings take different values.
An example of the evolution of correlation lengths and velocities is shown in Fig.~\ref{fig:scaling}. We find that the scaling solution is a good approximation in the radiation era ($t<10^{12}$~s) except for small fluctuations caused by the evolving number of relativistic degrees of freedom in the plasma.

\begin{table}[t!]
\caption{\label{Tab:super_string_chopping} The chopping efficiency $\tilde{c}_i$ and cross-interaction efficiency $\tilde{d}_{ij}^k$ in the radiation era for volume suppression factors $w=1$ and $w=0.1$~\cite{Pourtsidou:2010gu}.} 
    \begin{center}
    \begin{tabular}{ c c c c c c c } 
    \multirow{2}{*}{$g_s$}
                &&    \multicolumn{2}{c}{$w=1$} && \multicolumn{2}{c}{$w=0.1$}  \\ \cline{3-4} \cline{6-7}
         && $(\tilde{c}_1, ~\tilde{c}_2, ~\tilde{c}_3)$ & $(\tilde{d}_{12}^3, ~\tilde{d}_{13}^2, ~\tilde{d}_{23}^1)$ && $(\tilde{c}_1, ~\tilde{c}_2, ~\tilde{c}_3)$ & $(\tilde{d}_{12}^3, ~\tilde{d}_{13}^2, ~\tilde{d}_{23}^1)$ \\ \hline\hline  
         0.04&& (0.02, 0.13, 0.13) & (0.05, 0.08, 0.55) && (0.01, 0.13, 0.13) & (0.05, 0.07, 0.55) \\\hline
          0.1 && (0.03, 0.16, 0.16) & (0.04, 0.11, 0.62) && (0.02, 0.16, 0.16) & (0.04, 0.10, 0.62) \\\hline
          0.2 && (0.05, 0.19, 0.19) & (0.03, 0.14, 0.63) && (0.02, 0.19, 0.19) & (0.03, 0.13, 0.63) \\\hline
          0.3 && (0.07, 0.20, 0.20) & (0.03, 0.16, 0.61) && (0.03, 0.20, 0.20) & (0.02, 0.14, 0.61) \\\hline
          0.5 && (0.10, 0.21, 0.21) & (0.02, 0.21, 0.54) && (0.05, 0.20, 0.21) & (0.01, 0.15, 0.54) \\\hline
          0.7 && (0.12, 0.22, 0.22) & (0.02, 0.26, 0.49) && (0.06, 0.15, 0.22) & (0.01, 0.17, 0.39) \\\hline
          0.9 && (0.15, 0.22, 0.22) & (0.02, 0.31, 0.45) && (0.07, 0.12, 0.21) & (0.01, 0.20, 0.31) \\\hline
\end{tabular} 
    \end{center} 
\end{table}

 From Eq.~\eqref{eq:loop_production}, it is expected that a smaller correlation length leads to a denser population of loops, i.e., $n_i(l,t) \propto \xi_i^{-3}$. 
Thus, we write the number density distribution function for type-$i$ string loops as
\begin{eqnarray} \label{eq:number_density_super}
t^4 n_i(l,t)  = {\cal F} \frac{\tilde{c}_iv_{i,\star}}{\gamma_{v_i,\star} \xi_{i,\star}^3} \, 
 \frac{0.32}{(1 + t_{i,\star} \, \dot{\xi}_{i, \star}/\xi_{i,\star} )(l/t + \Gamma G \mu_i)^{\frac52}} \,,
\end{eqnarray}
where the time at loop production is
\begin{eqnarray}
    t_{i,\star} \simeq \frac{t}{\alpha_L \xi_{i,{\rm r}}} (l/t + \Gamma G \mu_i)\,.
\end{eqnarray}

\subsubsection{Scaling solution}

\begin{figure}[t!]
\centering
\includegraphics[width=.75\textwidth]{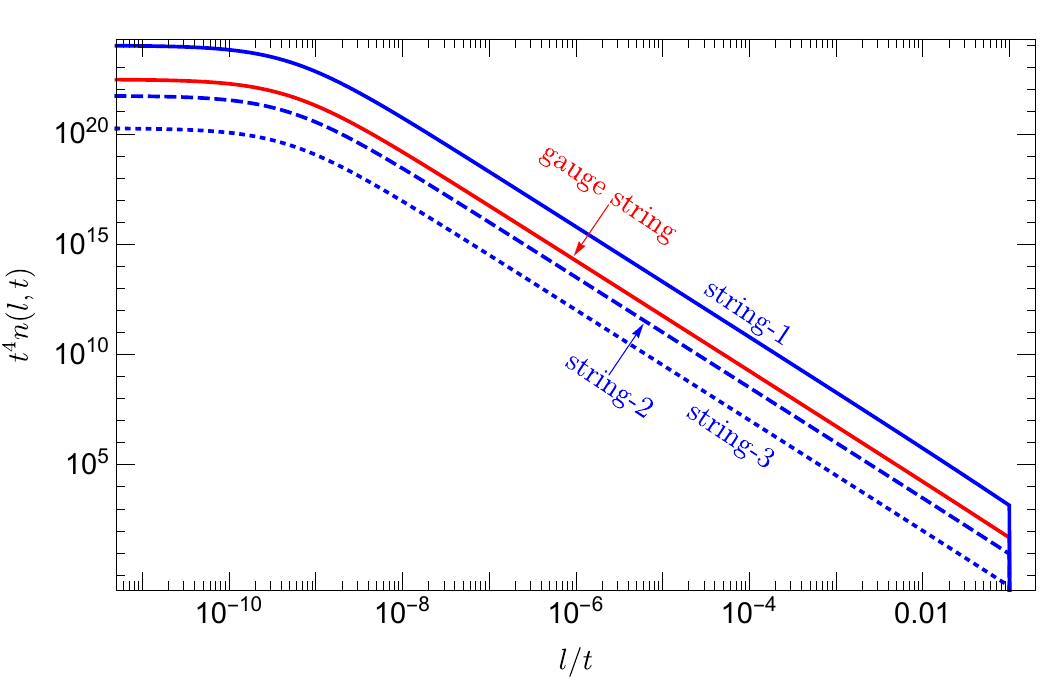}
\caption{Loop number density distribution $t^4 n_i(l,t)$ for three kinds of superstrings (in blue) at redshift $z= 10^9$ for $G\mu_1=10^{-11}$, $w=1$ and $g_s=0.1$. The distribution function for gauge strings with $G\mu=10^{-11}$ is shown in red for comparison.}
\label{fig:loop_density_super}
\end{figure}

We relate the number density of loops in superstring networks with $\xi_i$ and $v_i$ not fixed at $\xi_{\rm r}$ and $v_{\rm r}$ to that in gauge string networks in the scaling regime. In a superstring network,
\begin{eqnarray}
t^4 n_i(l,t)|_{\rm scaling} &=& \frac{0.18 }{(l/t + \Gamma G \mu_i)^{5/2}}  \times N_i \,,
\end{eqnarray}
where the first factor on the right hand side is identical to the loop number density in gauge string networks and $N_i$ is an enhancement/suppression factor for string-$i$ in the superstring network:
\begin{eqnarray} \label{eq:enhancement}
N_i= \frac{\tilde{c}_i}{\tilde{c}} \; \frac{v_i / \gamma_{v_i}}{v_{\rm r} / \gamma_{v_{\rm r}}} \; \left(\frac{\xi_{\rm r}}{\xi_i}\right)^3 
\simeq 0.04\; P_{ii}^{\frac13} \times \frac{v_i}{\gamma_{v_i} \xi_i^3} \,.
\end{eqnarray}
 It is noteworthy that the effect of the chopping and cross-interaction efficiencies is captured by normalizing the loop number density in gauge string networks by an overall factor $N_i$.
See Table~\ref{Tab:intercommunication} for values of $N_i$ for the benchmark points in Table~\ref{Tab:super_string_chopping}. The loop number densities for the three types of strings  at redshift $z=10^9$ for $g_s = 0.1$ and $w=1$ are shown in Fig.~\ref{fig:loop_density_super}. 
Since the correlation length for F-strings is usually much smaller than for gauge strings, the loop number density for superstrings is much larger than for gauge strings. 

We also checked the dependence of the loop number density on the intercommutation probability by setting $\tilde{d}_{jk}^i=0$. As $\xi_i\propto \tilde{c}_i \propto P_{ii}^{\frac13}$, a smaller intercommutation probability leaves a denser string network, where the dependence of $v_i$ on $\tilde{c}_i$ is much weaker and can be neglected. The loop number density is then enhanced due to the denser string network, but in combination with a smaller chopping efficiency we find, $n_i(l,t) \propto \tilde{c}_i /\xi_i^3 \propto \tilde{c}_i^{-2} \propto P_{ii}^{-\frac23}$.  
Numerically we have checked that, for F-strings, the enhancement due to the small intercommutation probability is given by $N_1 \sim {\cal O}(1) P_{11}^{-\frac23}$.

\begin{table}[t!]
\caption{\label{Tab:intercommunication}  
$N_i$ for the benchmark points in Table~\ref{Tab:super_string_chopping}.} 
\begin{center}
\begin{tabular}{ c c c c c c c c } 
\multirow{2}{*}{$g_s$}     &  \multicolumn{3}{c}{$w=1$}  && \multicolumn{3}{c}{$w=0.1$} \\\cline{2-4} \cline{6-8}
& $N_1$ & $N_2$ & $N_3$ && $N_1$ & $N_2$ & $N_3$ \\\hline\hline
 0.04 & 71.3 & 0.116 & 0.0543 && 238 & 0.101 & 0.0602 \\
 0.1 & 33.8 & 0.180 & 0.0336 && 73.6 & 0.156 & 0.0365 \\
 0.2 & 12.9 & 0.332 & 0.0220 && 75.4 & 0.259 & 0.0264 \\
 0.3 & 6.81 & 0.412 & 0.0188 && 35.0 & 0.422 & 0.0184 \\
 0.5 & 3.62 & 0.703 & 0.00975 && 13.5 & 0.789 & 0.00899 \\
 0.7 & 2.58 & 0.752 & 0.00814 && 9.29 & 1.38 & 0.0142 \\
 0.9 & 1.74 & 0.825 & 0.00669 && 6.69 & 2.17 & 0.0169 \\\hline
\end{tabular} 
    \end{center} 
\end{table}

\subsection{GWs from superstring loops}

The total GW spectrum is a sum of the contributions from the three types of strings, i.e., 
\begin{eqnarray}
\Omega_{\rm GW} = \sum_{i=1,2,3} \Omega_{i,\rm GW}\,,\quad
{d \rho_{\rm GW}\over df} = \sum_{i=1,2,3} {d\rho_{i,\rm GW} \over df}\,,
\end{eqnarray}
where
\begin{eqnarray}
{d \rho_{i,\rm GW}(t_0, f) \over df }= \int_0^{t_0} \frac{dt}{(1+z)^4} P_{i, \rm GW} (t,f') \frac{\partial f'}{\partial f}\,,
\end{eqnarray}
with $P_{i,\rm GW} (t,f')$ the GW power at frequency $f'$ from type-$i$ loops radiating at time $t$. $P_{i,\rm GW}$ depends on the string tension $\mu_i$ and loop number density $n_i(l,t)$ for type-$i$ strings. In analogy with the gauge string network, the GW energy density is 
\begin{eqnarray}
{d \rho_{\rm GW} (t_0,f)\over df} = \sum_{i=1,2,3} G\mu_i^2 \;\sum_{k=1}^{\infty} C_{i,k} P_{k}\,,
\end{eqnarray}
where 
\begin{eqnarray}
C_{i,k} = {\frac{2k}{f^2}} \int_0^{\infty} \frac{dz}{H(z)(1+z)^6} n_i\bigg(\frac{2k}{(1+z)f}, t(z)\bigg)\,.
\end{eqnarray}

\begin{figure}[t!]
\centering
\includegraphics[height=.4\textwidth]{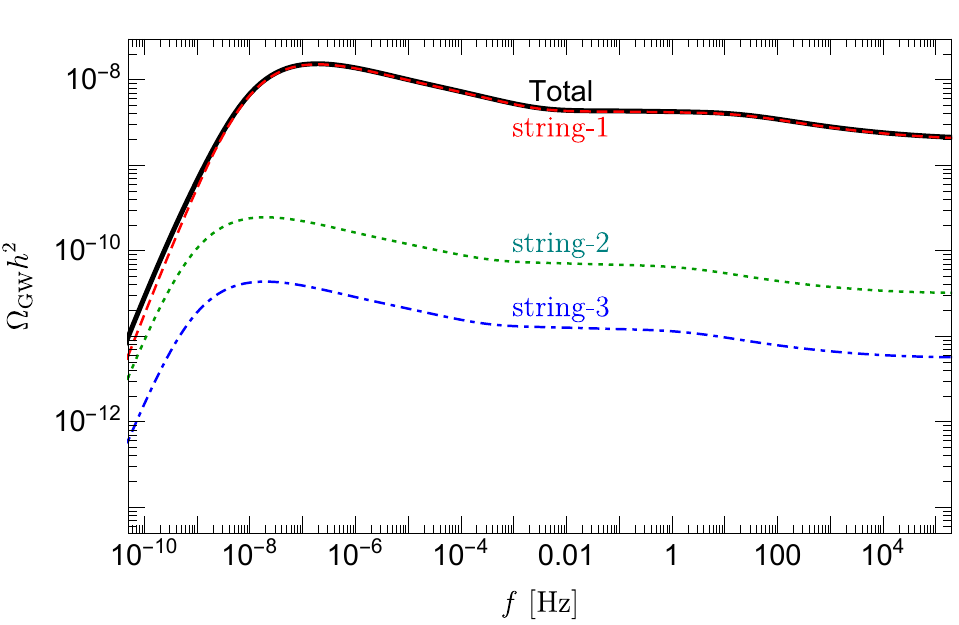}
\includegraphics[height=.4\textwidth]{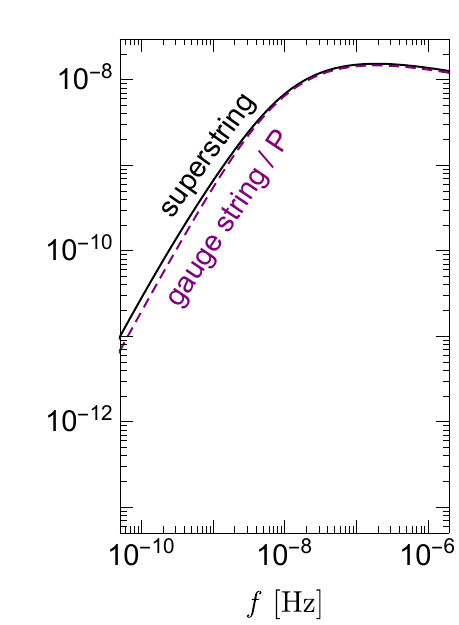}\caption{Left panel: Total GW spectrum from a superstring network with its individual components shown separately. We set $G\mu_1 = 10^{-11}$, $w=1$ and $g_s=0.1$. The chopping efficiency $\tilde{c}_i$ and cross-interaction efficiency $\tilde{d}_{ij}^k$ can be found in Table~\ref{Tab:super_string_chopping}. 
Right panel: The GW spectrum in the PTA band obtained by scaling the spectrum for a gauge string network by a free parameter $1/P$. The value of $1/P$ is chosen to match the total spectrum in the left panel in the high frequency band.}  
\label{fig:GW_superstring_1}
\end{figure}

An example of the superstring  GW spectrum is shown in Fig.~\ref{fig:GW_superstring_1}. 
The components of the total GW spectrum for different types of strings are shown separately in the left panel. The contribution from F-strings dominates because its small correlation length (shown in Fig.~\ref{fig:scaling}),
results in a denser network and correspondingly higher density of F-string loops.
However, other string types contribute nonnegligibly in the nHz band, and affect the shape of the GW spectrum. This is because D- and FD-strings have larger tension than F-strings, which extends the plateau of their spectrum to lower frequencies. The right panel shows a comparison of the GW spectrum in the PTA band  obtained by simply scaling the gauge string GW spectrum by $1/P$, with the total spectrum in the left panel. 

\begin{figure}[t!]
\centering
\includegraphics[width=.85\textwidth]{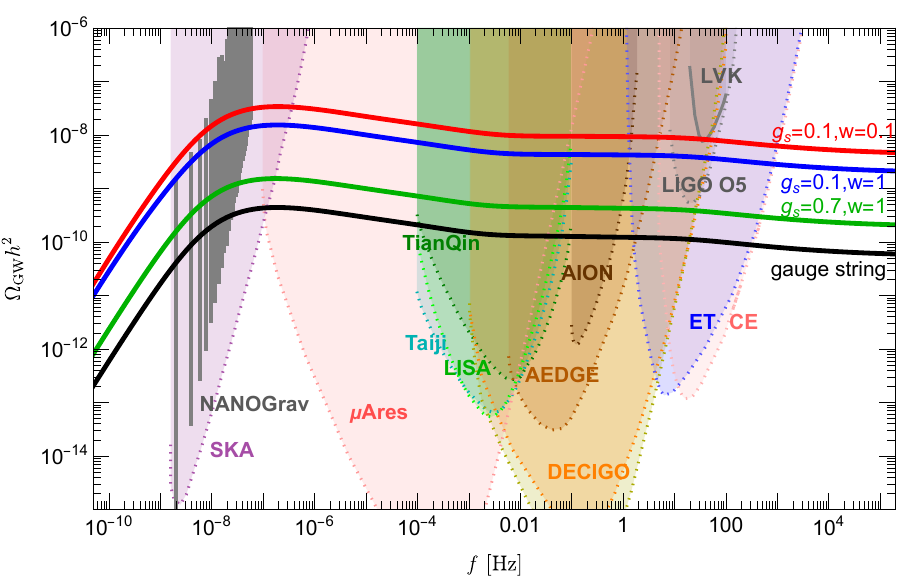}
\caption{A comparison of the GW spectra from cosmic superstrings and gauge strings. $G\mu_1$ for superstrings and $G\mu$ for gauge strings are both assumed to be $10^{-11}$.
The chopping efficiency $\tilde{c}_i$ and cross-interaction efficiency $\tilde{d}_{ij}^k$ can be found in Table~\ref{Tab:super_string_chopping}.
}
\label{fig:GW_superstring_2}
\end{figure}

In Fig.~\ref{fig:GW_superstring_2}, we compare the GW spectra from
cosmic superstring and gauge string networks. In the superstring case, a few examples with $(g_s, w) = (0.1, 0.1), (0.1, 1)$ and $(0.7, 1)$ are shown. Superstrings give a much stronger signal than gauge strings, and $w=0.1$ gives a stronger signal than $w=1$ since the intercommutation probability is smaller for smaller $w$. As expected from Table~\ref{Tab:intercommunication}, 
the GW signal for a given value of $w$ is stronger for smaller $g_s$.

%

 The analytical expressions for $\Omega_{\rm GW}(f)h^2$ in Section~\ref{sec:GW_formulism} are easily to generalized to the superstring case. 
The power-law approximation in the nHz band is given by
\begin{eqnarray}
 \Omega_{\rm GW}(f)h^2 |_{\text{power-law, PTA}} \simeq 1.96 \times 10^{-9}  \left( \frac{f}{f_{\rm yr}} \right)^{\frac32} \,\sum_i N_i \left( \frac{G\mu_i}{10^{-11}} \right)^2\,.
\end{eqnarray}
In the high frequency band, 
\begin{eqnarray}
 \Omega_{\rm GW}(f) h^2 |_{\rm laser} 
    &\simeq& 4.78 \times 10^{-5}\times \mathcal{C} \sum_i N_i \sqrt{G\mu_i} \,.
\end{eqnarray}
The power-law approximation in Eq.~(\ref{eq:Omegahsq_laser}) applies with the amplitude modified by the enhancement factors $N_i$:
\begin{eqnarray}
    A_{\rm laser} = 8.65 \cdot 10^{-16}  \times \sqrt{{\cal C}} \Bigg( \sum_{i}N_i \sqrt{\frac{G\mu_i}{10^{-11}}} \Bigg)^{\frac12} \,.
\end{eqnarray}

\section{Testing string networks with data}
\label{sec4}
\begin{figure}[t!]
\centering
\includegraphics[width=.48\textwidth]{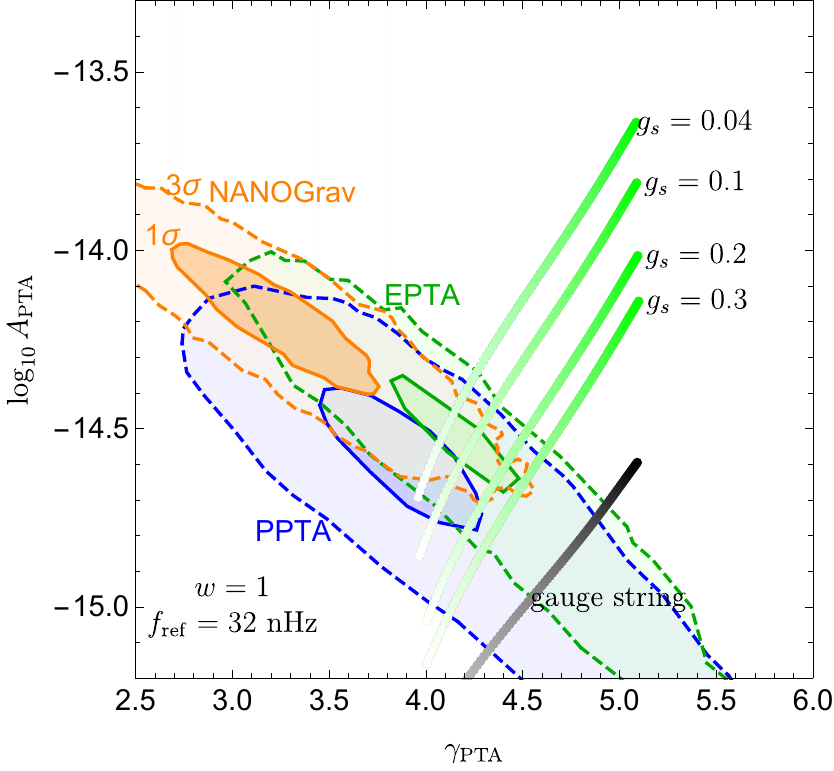}
\includegraphics[width=.48\textwidth]{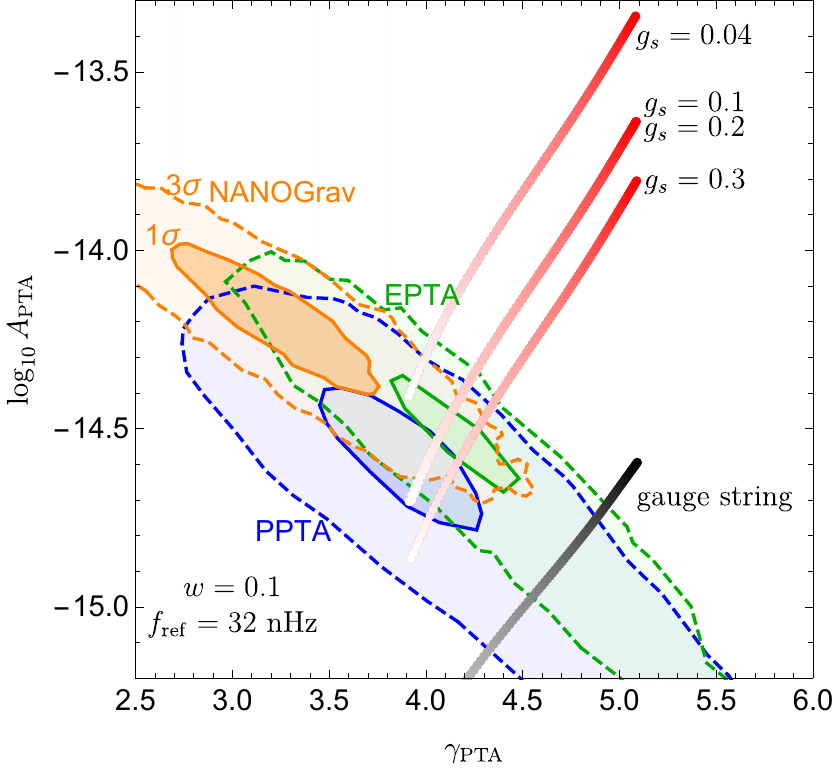}
\includegraphics[width=.48\textwidth]{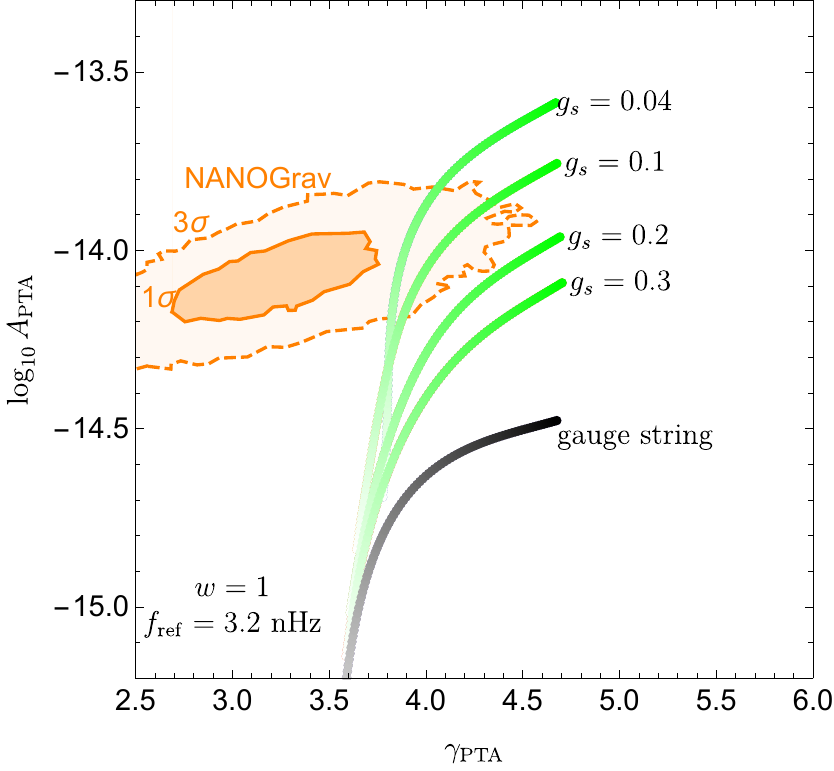}
\includegraphics[width=.48\textwidth]{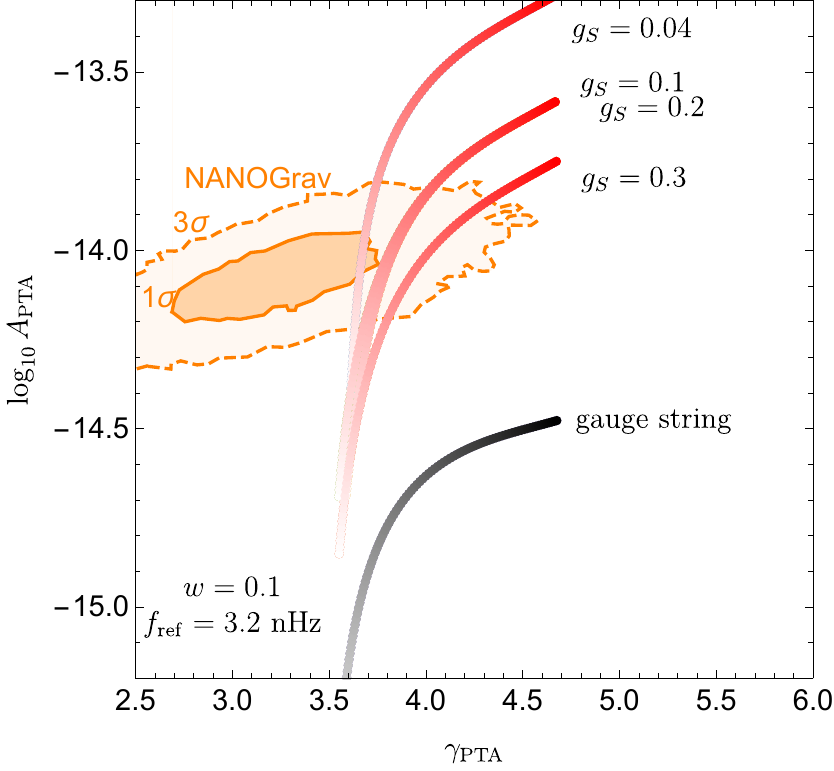}
\caption{Dependence of the parameters in the power-law approximation of the GW spectrum on $G\mu_1$ for several values of $g_s$. $G\mu_1$ is varied from $10^{-12}$ (lighter color) to $10^{-10}$ (darker color). 
The volume suppression factor is fixed at $w = 1$ (left panels) and 0.1 (right panels). The reference frequency is $f_{\rm ref} = 32$~nHz (upper panels) and $3.2$~nHz (lower panels). The gauge string case, as a comparison, is shown in black with $G\mu$ varied from $10^{-12}$ (lighter) to $10^{-10}$ (darker). In the right panels, the curves for $g_s=0.1$ and $g_s=0.2$  overlap because they have the same values of $\tilde{c}_i$, as can be seen in Table~\ref{Tab:super_string_chopping}. The $1\sigma$ and $3\sigma$ allowed regions from NANOGrav-15, EPTA DR2full, and PPTA data are also shown.}
\label{fig:A_gamma}
\end{figure}

\begin{figure}[th]
\centering
\includegraphics[width=.7\textwidth]{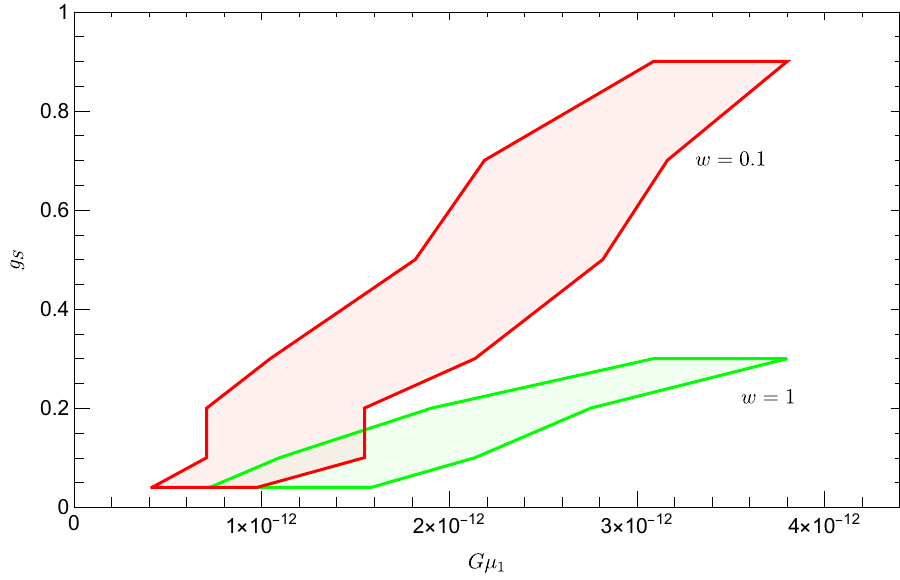}
\caption{$3\sigma$ region in the $(G\mu_1,g_s)$ plane allowed by NANOGrav-15 for $f_{\rm ref} = 32$~nHz. The boundaries of the regions are jagged because of the small number of benchmark points and the limited precision of the chopping efficiencies in Table~\ref{Tab:super_string_chopping}.}
\label{fig:gSGmu}
\end{figure}

The 15 year NANOGrav dataset, NANOGrav-15~\cite{NANOGrav:2023gor}, and the EPTA DR2full~\cite{Antoniadis:2023ott} and PPTA~\cite{Reardon:2023gzh} datasets have been analyzed using a power-law approximation. We first check the consistency of the power-law approximated GW spectrum with PTA data. Given a reference frequency $f_{\rm ref} =32$~nHz , we show in the upper panel of Fig.~\ref{fig:A_gamma} how $A_{\rm PTA}$ and $\gamma_{\rm PTA}$ depend on $G\mu_1$ for $w=1$ (left panel) and $w=0.1$ (right panel). For a fixed value of $g_s$ the variation of $G\mu_1$ from $10^{-12}$ to $10^{-10}$ is depicted with a darkening shade of green and red, respectively. Note that different values of $g_s$ yield different input values of $\tilde{c}_i$ and $\tilde{d}_{ij}^k$, as in Table~\ref{Tab:super_string_chopping}.
For $w=0.1$, the $g_s=0.1$ and $g_s=0.2$ curves overlap because the values of $\tilde{c}_i$ and $\tilde{d}_{ij}^k$ are almost the same, which leads to very similar values of $N_1$ in Table~\ref{Tab:intercommunication}. 
The ($f_{\rm ref}$-dependent) $1\sigma$ and $3\sigma$ C.L. regions in the ($\gamma_{\rm PTA},A_{\rm PTA}$) plane favored by NANOGrav-15, EPTA and PPTA data are also shown. As has been already pointed out, a gauge string network is not compatible with NANOGrav-15 data at $3\sigma$~\cite{NANOGrav:2023hvm,Ellis:2023tsl}. However, superstring networks with $w=1$ are compatible at $3\sigma$, and the case with $w=0.1$ is even more compatible with the data. As can be seen from the figure, EPTA and PPTA data are in mild tension with the NANOGrav-15 data, and do not exclude a gauge string network at $3\sigma$. However, cosmic superstrings are consistent with these data to within $1\sigma$.  
NANOGrav has also provided the preferred region in the ($\gamma_{\rm PTA},A_{\rm PTA}$) plane for $f_{\rm ref} = 3.2$~nHz. From the lower panels of Fig.~\ref{fig:A_gamma}, we see that while some superstring networks with $w=1$ are excluded at $3\sigma$ unless $g_s< 0.2$, networks with $w=0.1$ are consistent with the data. Gauge string networks are even less compatible than for $f_{\rm ref} = 32$~nHz.

For the values of $g_s$ in Table~\ref{Tab:super_string_chopping}, we identify the lower and upper bounds of $G\mu_1$ that are consistent with the $3\sigma$ region allowed by NANOGrav-15 with $f_{\rm ref}=32$~nHz. We then obtain the $3\sigma$ region in the $(G\mu_1, g_s)$ plane shown in Fig.~\ref{fig:gSGmu}. We have included this figure for illustrative purposes only since the small number of benchmark points and limited precision of the chopping coefficients in Table~\ref{Tab:super_string_chopping} yields quite jagged boundaries. 

\begin{figure}[t]
\centering
\includegraphics[width=.75\textwidth]{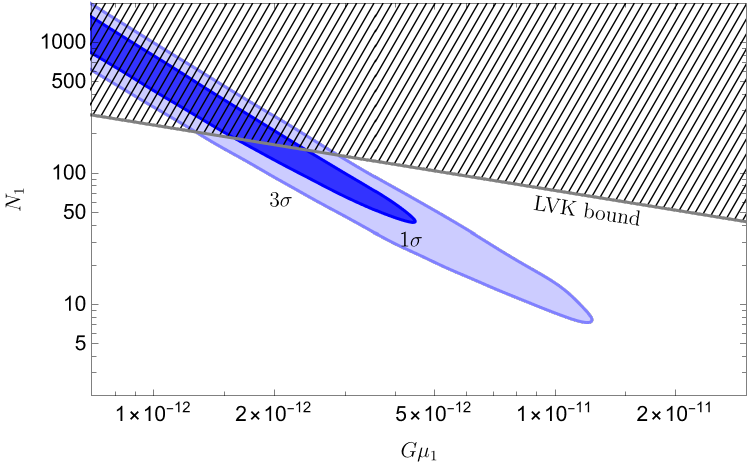}
\caption{$1\sigma$ and $3\sigma$ regions in the $(G\mu_1,N_1)$ plane allowed by NANOGrav-15 data.
 The contributions from D-strings and FD-strings are negligible since their number densities are highly suppressed. The  LVK 95\% CL upper bound excludes most of the $1\sigma$ allowed region.}
\label{fig:NGmu}
\end{figure}

The {\tt PTArcade} software provides an accessible way to perform Bayesian analyses of new physics signals with PTA data~\cite{Mitridate:2023oar,Lamb:2023jls}. 
It enables direct analysis of a given GW spectrum without resorting to the power-law approximation.
In the superstring case, constraints on the string tensions $G\mu_i$ and the enhancement factors $N_i$, defined in Eq.~\eqref{eq:enhancement}, can be obtained. As a simplification, we neglect the GW contributions from string-2 and string-3 since their number densities are subdominant to string-1's. Then, the GW spectrum is determined by two parameters, $G\mu_1$ and $N_1$. The $1\sigma$ and $3\sigma$ regions allowed by NANOGrav-15 data are shown in Fig.~\ref{fig:NGmu}.
The factor of $\sim 2$ difference in the maximum value of $G\mu_1$
allowed at $3\sigma$ in Figs.~\ref{fig:gSGmu} and~\ref{fig:NGmu} is due to the power-law approximation used in the former.
Despite this small gap, both show that $G\mu_1 \lesssim 10^{-11}$ is preferred by NANOGrav-15 data. The LIGO-Virgo-KARGRA (LVK) 95\%~CL upper bound at $\sim 25$~Hz, $\Omega_{\rm GW}(f)h^2 \lesssim 7.8 \times 10^{-9}$~\cite{Abbott:2021xxi}, disfavors a large enhancement factor $N_1 \gtrsim 200$ and a small string tension $G\mu_1 \lesssim 1.5\times 10^{-12}$; see Fig.~\ref{fig:NGmu}. A region defined by $N_1 \simeq {\cal O}(10\text{-}100)$ and $G\mu_1 \simeq {\cal O}(10^{-11})$ is compatible with PTA measurements and the LVK constraint. 

We note that for current data, the GW spectrum from superstring networks is adequately approximated by simply scaling the GW spectrum for gauge string networks by the inverse of the intercommutation probability; this is evident from the right panel of Fig.~\ref{fig:GW_superstring_1}. However, as can be seen from Fig.~\ref{fig:A_gamma}, the values of $(g_s, w)$ play an important role for a more accurate assessment. 
As PTA data get more precise, modifications in the shape of the GW spectrum due to the different superstring types will need to be taken into account.

\begin{figure}[t!]
\centering
\includegraphics[height=.42\textwidth]{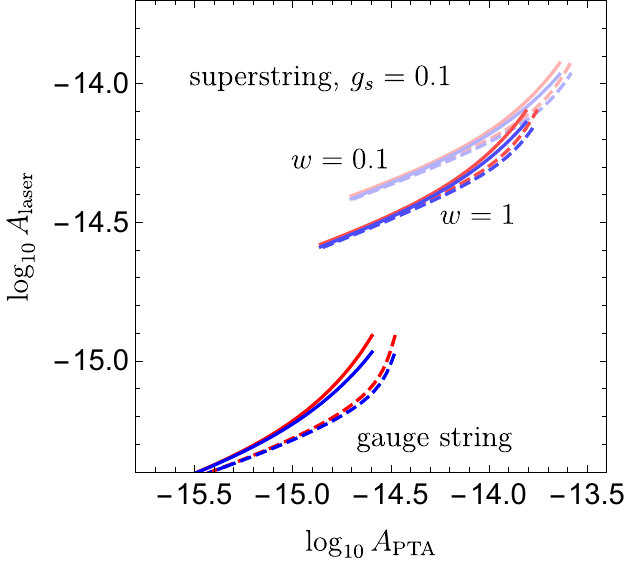}
\includegraphics[height=.42\textwidth]{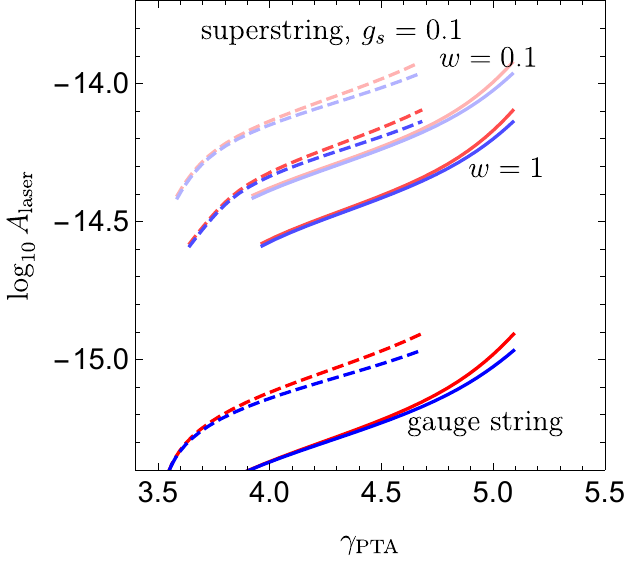}
\caption{Correlation between $A_{\rm PTA}$ and $A_{\rm laser}$ (left panel) and that between $\gamma_{\rm PTA}$ and $A_{\rm laser}$ (right panel) for gauge strings and for superstrings with $g_s= 0.1$ and $w=0.1,1$. $G\mu$ and $G\mu_1$ are varied from $10^{-12}$ to $10^{-10}$. The reference frequencies for PTA are $f_{\rm ref} =$ 3.2~nHz (dashed) and 32~nHz (solid), and for laser interferometers, $f_{\rm ref} =$ 0.01~Hz (red) and 10~Hz (blue).
}
\label{fig:PTA_LISA}
\end{figure} 

We finally study correlations between measurements at PTAs and laser interferometers. 
In Fig.~\ref{fig:PTA_LISA}, we show correlations in the power-law parameters with $G\mu$ and $G\mu_1$ varying from $10^{-12}$ to $10^{-10}$.  
Since the GW spectrum for gauge string networks depends mainly on a single parameter $G\mu$, tight correlations between $A_{\rm PTA}$ and $A_{\rm laser}$, and between $\gamma_{\rm PTA}$ and $A_{\rm laser}$ are evident. On the other hand, the GW spectrum
for superstring networks depends on $G\mu_i$, $g_s$ and $w$, which considerably enlarges the spread of possible signals, and weakens correlations. This can be seen from the wide separation between the superstring curves in Fig.~\ref{fig:PTA_LISA}.
The  clear separation between the dashed and solid curves in the right panel indicates that the power-law approximation is not a good description of the GW spectrum in the PTA regime. In the superstring case,  we take $g_s=0.1$ and $w=0.1,1$. Once these parameters are fixed, correlations in the power-law parameters look similar to that in the gauge string case. However, the enhancement in the loop number density shifts the curves to a region not accessible by gauge string networks. 
The large gap between the curves along the $A_{\rm laser}$ axis confirms that the signal amplitude at high frequencies is an obvious discriminator of superstring and gauge string networks.

\begin{figure}[t]
\centering
\includegraphics[height=.45\textwidth]{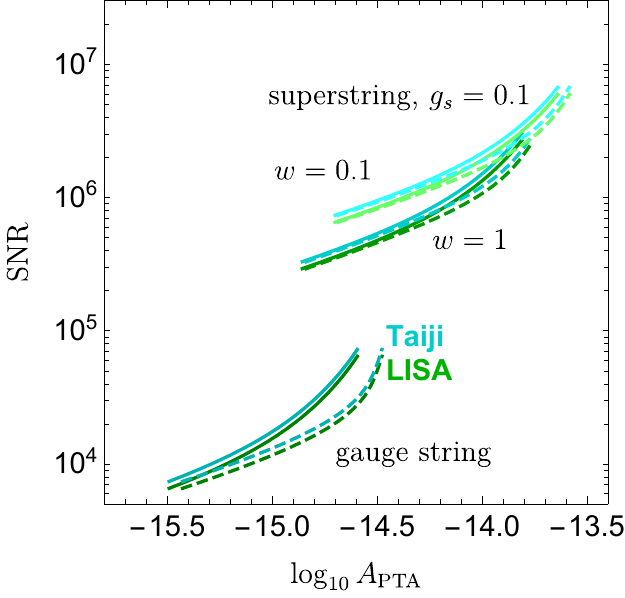}
\includegraphics[height=.45\textwidth]{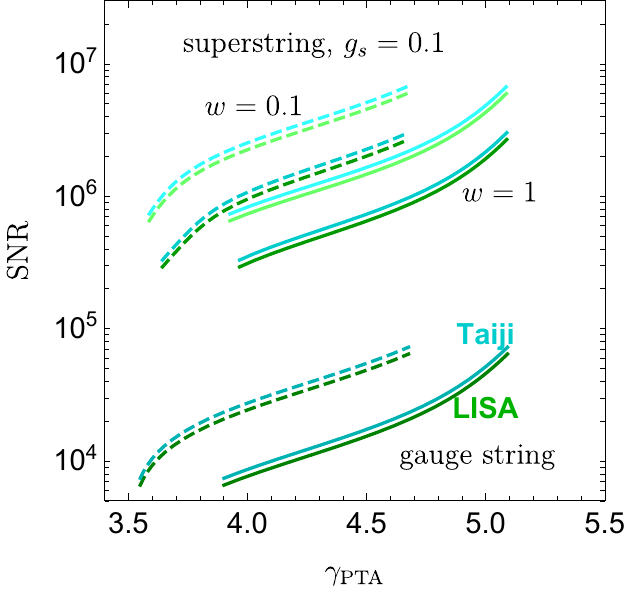}
\caption{Signal-to-noise ratio of GWs at LISA and Taiji for a 4-year observation time. The reference frequencies for
PTA are $f_{\rm ref} = 3.2$ nHz (dashed) and 32 nHz (solid). 
$G\mu$ and $G\mu_1$ are varied from $10^{-12}$ to $10^{-10}$, and
$g_s=0.1$ and $w=0.1,1$ are assumed for superstrings. } 
\label{fig:PTA_LISA_SNR}
\end{figure}

The figure of merit for typical space-based laser interferometers like LISA and Taiji, is the signal-to-noise ratio (SNR), which is defined by
\begin{eqnarray}
    {\rm SNR} = \sqrt{T \int_0^{\infty} df \frac{\Omega_{\rm GW}^2(f)}{\Omega_{n}^2(f)}} \,.
\end{eqnarray}
Here, $\Omega_{n}(f) = \frac{2\pi^2}{H_0^2} f^3 S_n(f)$ normalizes the noise power spectral density $S_n(f)$ to an equivalent GW energy density. We take $\Omega_{n}(f)$ for LISA and Taiji from Refs.~\cite{Robson:2018ifk} and \cite{Guo:2018npi}, respectively, and consider a 4-year observation time, $T = 1.26 \times 10^8$.  Since the GW spectrum is flat at high frequencies, we fix $\Omega_{\rm GW}(f)$ at a typical reference frequency, e.g., $f_{\rm ref} = 0.01$~Hz. Then 
\begin{eqnarray}
    {\rm SNR} \simeq \Omega_{\rm GW}(f_{\rm ref}) \times \sqrt{T \int_0^{\infty} df\ \Omega_{n}^{-2}(f)} \,
\end{eqnarray}
is the GW energy density up to a normalization factor fixed by the experimental setup. In Fig.~\ref{fig:PTA_LISA_SNR}, we vary $G\mu$ and $G\mu_1$ and plot SNR versus the PTA power-law parameters. The qualitative conclusions drawn from Fig.~\ref{fig:PTA_LISA} are quantified in terms of SNR at LISA and Taiji in this figure.

\section{Summary}
\label{sec5}

We studied GW signals from the decay of loops in cosmic superstring and gauge string networks. In superstring networks with F-, D- and FD-strings, we find that F-strings provide the strongest GW signal at laser interferometers, while D-strings and FD-strings affect the GW spectrum at PTAs. For networks with equal gauge string and F-string tensions, the superstring network gives a significantly stronger GW signal because the number density of loops scales inversely with the intercommutation probability. 

We considered effects of the evolution of the number of relativistic degrees of freedom on the GW spectrum. The spectrum deviates from the scaling solution at the level of $\sim 10\%$ unless a large number of non-Standard Model particles contribute to the thermal bath.
We showed that if this evolution is properly modeled, then the scaling approximation gives a GW spectrum that matches the more precise nonscaling solution, which accounts for the variation of the correlation length as the Universe expands, for all frequencies relevant to PTAs and laser interferometers. We also provided analytic expressions for the GW
spectrum from superstrings and gauge strings that are valid for these frequencies. Equations~\eqref{eq:Omega_hsq_PTA_approx} and \eqref{eq:Omega_hsq_laser} are simple enough to facilitate rapid calculations of GW spectra from string networks.

We analyzed recent data from NANOGrav-15, EPTA and PPTA, and showed that since EPTA and PPTA are in mild tension with NANOGrav data, gauge string networks are not as strongly excluded as 
portrayed. Also, only superstring networks in which the strings
evolve in only a small fraction of the higher-dimensional space are compatible with 3.2~nHz from NANOGrav. 
In the superstring case, the GW spectrum is determined mainly by the tension $\mu_1$ and the loop enhancement factor $N_1$ of F-strings. Parameter values that are  favored by NANOGrav-15 data and that are consistent with the LVK bound are $G\mu_1 \simeq 10^{-11}$ and $N\simeq {\cal O}(10-100)$.
We also studied correlations between GW signals at PTAs and laser interferometers. We find that correlations between the power-law parameters describing the GW spectrum are similar for superstring and gauge string networks for fixed values of $g_s$ and $w$. However, the power-law parameters for the two kinds of networks occupy distinct regions; see Fig.~\ref{fig:PTA_LISA}.

\section*{Acknowledgements}
We thank X.-H.~Hu for drawing our attention to typographical errors.
Y.L.Z thanks H.K.~Guo, Z.K.~Guo and C.~Liu for useful discussions.
D.M. is supported in part by the U.S. Department of Energy under Grant No.~de-sc0010504. 
Y.L.Z. is supported by the National Natural Science Foundation of China (NSFC) under Grant Nos.~12205064, 12347103, and Zhejiang Provincial Natural Science Foundation of China under Grant No. LDQ24A050002.


\begin{thebibliography}{99}

\bibitem{Kibble:1976sj}
T.~W.~B.~Kibble,
J. Phys. A \textbf{9} (1976), 1387-1398
doi:10.1088/0305-4470/9/8/029

\bibitem{Jones:2002cv}
N.~T.~Jones, H.~Stoica and S.~H.~H.~Tye,
JHEP \textbf{07} (2002), 051
doi:10.1088/1126-6708/2002/07/051
[arXiv:hep-th/0203163 [hep-th]].

\bibitem{Sarangi:2002yt}
S.~Sarangi and S.~H.~H.~Tye,
Phys. Lett. B \textbf{536} (2002), 185-192
doi:10.1016/S0370-2693(02)01824-5
[arXiv:hep-th/0204074 [hep-th]].


\bibitem{Dvali:2003zj}
G.~Dvali and A.~Vilenkin,
JCAP \textbf{03} (2004), 010
doi:10.1088/1475-7516/2004/03/010
[arXiv:hep-th/0312007 [hep-th]].

\bibitem{Copeland:2003bj}
E.~J.~Copeland, R.~C.~Myers and J.~Polchinski,
JHEP \textbf{06} (2004), 013
doi:10.1088/1126-6708/2004/06/013
[arXiv:hep-th/0312067 [hep-th]].

\bibitem{Jackson:2004zg}
M.~G.~Jackson, N.~T.~Jones and J.~Polchinski,
JHEP \textbf{10} (2005), 013
doi:10.1088/1126-6708/2005/10/013
[arXiv:hep-th/0405229 [hep-th]].

\bibitem{Albrecht:1984xv}
A.~Albrecht and N.~Turok,
Phys. Rev. Lett. \textbf{54} (1985), 1868-1871
doi:10.1103/PhysRevLett.54.1868

\bibitem{Bennett:1987vf}
D.~P.~Bennett and F.~R.~Bouchet,
Phys. Rev. Lett. \textbf{60} (1988), 257
doi:10.1103/PhysRevLett.60.257

\bibitem{Allen:1990tv}
B.~Allen and E.~P.~S.~Shellard,
Phys. Rev. Lett. \textbf{64} (1990), 119-122
doi:10.1103/PhysRevLett.64.119

\bibitem{Jeannerot:2003qv}
R.~Jeannerot, J.~Rocher and M.~Sakellariadou,
Phys. Rev. D \textbf{68} (2003), 103514
doi:10.1103/PhysRevD.68.103514
[arXiv:hep-ph/0308134 [hep-ph]].

\bibitem{NANOGrav:2023gor}
G.~Agazie \textit{et al.} [NANOGrav],
Astrophys. J. Lett. \textbf{951} (2023) no.1, L8
doi:10.3847/2041-8213/acdac6
[arXiv:2306.16213 [astro-ph.HE]].

\bibitem{Antoniadis:2023ott}
J.~Antoniadis \textit{et al.} [EPTA and InPTA:],
Astron. Astrophys. \textbf{678} (2023), A50
doi:10.1051/0004-6361/202346844
[arXiv:2306.16214 [astro-ph.HE]].

\bibitem{Reardon:2023gzh}
D.~J.~Reardon, A.~Zic, R.~M.~Shannon, G.~B.~Hobbs, M.~Bailes, V.~Di Marco, A.~Kapur, A.~F.~Rogers, E.~Thrane and J.~Askew, \textit{et al.}
Astrophys. J. Lett. \textbf{951} (2023) no.1, L6
doi:10.3847/2041-8213/acdd02
[arXiv:2306.16215 [astro-ph.HE]].

\bibitem{Xu:2023wog}
H.~Xu, S.~Chen, Y.~Guo, J.~Jiang, B.~Wang, J.~Xu, Z.~Xue, R.~N.~Caballero, J.~Yuan and Y.~Xu, \textit{et al.}
Res. Astron. Astrophys. \textbf{23} (2023) no.7, 075024
doi:10.1088/1674-4527/acdfa5
[arXiv:2306.16216 [astro-ph.HE]].

\bibitem{NANOGrav:2023hvm}
A.~Afzal \textit{et al.} [NANOGrav],
Astrophys. J. Lett. \textbf{951}, no.1, L11 (2023)
doi:10.3847/2041-8213/acdc91
[arXiv:2306.16219 [astro-ph.HE]].

\bibitem{Ellis:2023tsl}
J.~Ellis, M.~Lewicki, C.~Lin and V.~Vaskonen,
Phys. Rev. D \textbf{108}, no.10, 103511 (2023)
doi:10.1103/PhysRevD.108.103511
[arXiv:2306.17147 [astro-ph.CO]].


\bibitem{Ellis:2020ena}
J.~Ellis and M.~Lewicki,
Phys. Rev. Lett. \textbf{126} (2021) no.4, 041304
doi:10.1103/PhysRevLett.126.041304
[arXiv:2009.06555 [astro-ph.CO]].

\bibitem{Blasi:2020mfx}
S.~Blasi, V.~Brdar and K.~Schmitz,
Phys. Rev. Lett. \textbf{126}, no.4, 041305 (2021)
doi:10.1103/PhysRevLett.126.041305
[arXiv:2009.06607 [astro-ph.CO]].

\bibitem{Blanco-Pillado:2021ygr}
J.~J.~Blanco-Pillado, K.~D.~Olum and J.~M.~Wachter,
Phys. Rev. D \textbf{103} (2021) no.10, 103512
doi:10.1103/PhysRevD.103.103512
[arXiv:2102.08194 [astro-ph.CO]].

\bibitem{Audley:2017drz}
P.~Amaro-Seoane \textit{et al.} [LISA],
[arXiv:1702.00786 [astro-ph.IM]].

\bibitem{Guo:2018npi}
W.~H.~Ruan, Z.~K.~Guo, R.~G.~Cai and Y.~Z.~Zhang,
Int. J. Mod. Phys. A \textbf{35} (2020) no.17, 2050075
doi:10.1142/S0217751X2050075X
[arXiv:1807.09495 [gr-qc]].

\bibitem{Luo:2015ght}
J.~Luo \textit{et al.} [TianQin],
Class. Quant. Grav. \textbf{33} (2016) no.3, 035010
doi:10.1088/0264-9381/33/3/035010
[arXiv:1512.02076 [astro-ph.IM]].

\bibitem{Corbin:2005ny}
V.~Corbin and N.~J.~Cornish,
Class. Quant. Grav. \textbf{23} (2006), 2435-2446
doi:10.1088/0264-9381/23/7/014
[arXiv:gr-qc/0512039 [gr-qc]].

\bibitem{Seto:2001qf}
N.~Seto, S.~Kawamura and T.~Nakamura,
Phys. Rev. Lett. \textbf{87} (2001), 221103
doi:10.1103/PhysRevLett.87.221103
[arXiv:astro-ph/0108011 [astro-ph]].

\bibitem{Sesana:2019vho}
A.~Sesana, N.~Korsakova, M.~A.~Sedda, V.~Baibhav, E.~Barausse, S.~Barke, E.~Berti, M.~Bonetti, P.~R.~Capelo and C.~Caprini, \textit{et al.}
Exper. Astron. \textbf{51} (2021) no.3, 1333-1383
doi:10.1007/s10686-021-09709-9
[arXiv:1908.11391 [astro-ph.IM]].

\bibitem{Graham:2017pmn}
P.~W.~Graham \textit{et al.} [MAGIS],
[arXiv:1711.02225 [astro-ph.IM]].

\bibitem{Bertoldi:2019tck}
Y.~A.~El-Neaj \textit{et al.} [AEDGE],
EPJ Quant. Technol. \textbf{7} (2020), 6
doi:10.1140/epjqt/s40507-020-0080-0
[arXiv:1908.00802 [gr-qc]].

\bibitem{Badurina:2019hst}
L.~Badurina, E.~Bentine, D.~Blas, K.~Bongs, D.~Bortoletto, T.~Bowcock, K.~Bridges, W.~Bowden, O.~Buchmueller and C.~Burrage, \textit{et al.}
JCAP \textbf{05} (2020), 011
doi:10.1088/1475-7516/2020/05/011
[arXiv:1911.11755 [astro-ph.CO]].

\bibitem{Sathyaprakash:2012jk}
B.~Sathyaprakash, M.~Abernathy, F.~Acernese, P.~Ajith, B.~Allen, P.~Amaro-Seoane, N.~Andersson, S.~Aoudia, K.~Arun and P.~Astone, \textit{et al.}
Class. Quant. Grav. \textbf{29} (2012), 124013
[erratum: Class. Quant. Grav. \textbf{30} (2013), 079501]
doi:10.1088/0264-9381/29/12/124013
[arXiv:1206.0331 [gr-qc]].

\bibitem{Evans:2016mbw}
B.~P.~Abbott \textit{et al.} [LIGO Scientific],
Class. Quant. Grav. \textbf{34} (2017) no.4, 044001
doi:10.1088/1361-6382/aa51f4
[arXiv:1607.08697 [astro-ph.IM]].

\bibitem{Janssen:2014dka}
G.~Janssen, G.~Hobbs, M.~McLaughlin, C.~Bassa, A.~T.~Deller, M.~Kramer, K.~Lee, C.~Mingarelli, P.~Rosado and S.~Sanidas, \textit{et al.}
PoS \textbf{AASKA14} (2015), 037
doi:10.22323/1.215.0037
[arXiv:1501.00127 [astro-ph.IM]].

\bibitem{Belanger:2018ccd}
G.~B\'elanger, F.~Boudjema, A.~Goudelis, A.~Pukhov and B.~Zaldivar,
Comput. Phys. Commun. \textbf{231} (2018), 173-186
doi:10.1016/j.cpc.2018.04.027
[arXiv:1801.03509 [hep-ph]].

\bibitem{Blanco-Pillado:2017oxo}
J.~J.~Blanco-Pillado and K.~D.~Olum,
Phys. Rev. D \textbf{96} (2017) no.10, 104046
doi:10.1103/PhysRevD.96.104046
[arXiv:1709.02693 [astro-ph.CO]].

\bibitem{Cui:2018rwi}
Y.~Cui, M.~Lewicki, D.~E.~Morrissey and J.~D.~Wells,
JHEP \textbf{01} (2019), 081
doi:10.1007/JHEP01(2019)081
[arXiv:1808.08968 [hep-ph]].

\bibitem{Auclair:2019wcv}
P.~Auclair, J.~J.~Blanco-Pillado, D.~G.~Figueroa, A.~C.~Jenkins, M.~Lewicki, M.~Sakellariadou, S.~Sanidas, L.~Sousa, D.~A.~Steer and J.~M.~Wachter, \textit{et al.}
JCAP \textbf{04}, 034 (2020)
doi:10.1088/1475-7516/2020/04/034
[arXiv:1909.00819 [astro-ph.CO]].

\bibitem{LIGOScientific:2021nrg}
R.~Abbott \textit{et al.} [LIGO Scientific, Virgo and KAGRA],
Phys. Rev. Lett. \textbf{126} (2021) no.24, 241102
doi:10.1103/PhysRevLett.126.241102
[arXiv:2101.12248 [gr-qc]].

\bibitem{Martins:1995tg}
C.~J.~A.~P.~Martins and E.~P.~S.~Shellard,
Phys. Rev. D \textbf{53} (1996), 575-579
doi:10.1103/PhysRevD.53.R575
[arXiv:hep-ph/9507335 [hep-ph]].

\bibitem{Martins:1996jp}
C.~J.~A.~P.~Martins and E.~P.~S.~Shellard,
Phys. Rev. D \textbf{54} (1996), 2535-2556
doi:10.1103/PhysRevD.54.2535
[arXiv:hep-ph/9602271 [hep-ph]].

\bibitem{Martins:2000cs}
C.~J.~A.~P.~Martins and E.~P.~S.~Shellard,
Phys. Rev. D \textbf{65} (2002), 043514
doi:10.1103/PhysRevD.65.043514
[arXiv:hep-ph/0003298 [hep-ph]].

\bibitem{Kibble:1984hp}
T.~W.~B.~Kibble,
Nucl. Phys. B \textbf{252} (1985), 227
[erratum: Nucl. Phys. B \textbf{261} (1985), 750]
doi:10.1016/0550-3213(85)90596-6



\bibitem{Blanco-Pillado:2013qja}
J.~J.~Blanco-Pillado, K.~D.~Olum and B.~Shlaer,
Phys. Rev. D \textbf{89} (2014) no.2, 023512
doi:10.1103/PhysRevD.89.023512
[arXiv:1309.6637 [astro-ph.CO]].

\bibitem{Sanidas:2012ee}
S.~A.~Sanidas, R.~A.~Battye and B.~W.~Stappers,
Phys. Rev. D \textbf{85} (2012), 122003
doi:10.1103/PhysRevD.85.122003
[arXiv:1201.2419 [astro-ph.CO]].

\bibitem{Sousa:2013aaa}
L.~Sousa and P.~P.~Avelino,
Phys. Rev. D \textbf{88} (2013) no.2, 023516
doi:10.1103/PhysRevD.88.023516
[arXiv:1304.2445 [astro-ph.CO]].

\bibitem{Auclair:2022ylu}
P.~Auclair, S.~Blasi, V.~Brdar and K.~Schmitz,
JCAP \textbf{04} (2023), 009
doi:10.1088/1475-7516/2023/04/009
[arXiv:2207.03510 [astro-ph.CO]].

\bibitem{Planck:2018vyg}
N.~Aghanim \textit{et al.} [Planck],
Astron. Astrophys. \textbf{641}, A6 (2020)
[erratum: Astron. Astrophys. \textbf{652}, C4 (2021)]
doi:10.1051/0004-6361/201833910
[arXiv:1807.06209 [astro-ph.CO]].

\bibitem{Fu:2022lrn}
B.~Fu, S.~F.~King, L.~Marsili, S.~Pascoli, J.~Turner and Y.~L.~Zhou,
JHEP \textbf{11} (2022), 072
doi:10.1007/JHEP11(2022)072
[arXiv:2209.00021 [hep-ph]].

\bibitem{Pourtsidou:2010gu}
A.~Pourtsidou, A.~Avgoustidis, E.~J.~Copeland, L.~Pogosian and D.~A.~Steer,
Phys. Rev. D \textbf{83} (2011), 063525
doi:10.1103/PhysRevD.83.063525
[arXiv:1012.5014 [astro-ph.CO]].

\bibitem{Avgoustidis:2007aa}
A.~Avgoustidis and E.~P.~S.~Shellard,
Phys. Rev. D \textbf{78} (2008), 103510
[erratum: Phys. Rev. D \textbf{80} (2009), 129907]
doi:10.1103/PhysRevD.78.103510
[arXiv:0705.3395 [astro-ph]].

\bibitem{Copeland:2009ga}
E.~J.~Copeland and T.~W.~B.~Kibble,
Proc. Roy. Soc. Lond. A \textbf{466} (2010), 623-657
doi:10.1098/rspa.2009.0591
[arXiv:0911.1345 [hep-th]].

\bibitem{Avgoustidis:2009ke}
A.~Avgoustidis and E.~J.~Copeland,
Phys. Rev. D \textbf{81} (2010), 063517
doi:10.1103/PhysRevD.81.063517
[arXiv:0912.4004 [hep-ph]].

\bibitem{Sousa:2016ggw}
L.~Sousa and P.~P.~Avelino,
Phys. Rev. D \textbf{94} (2016) no.6, 063529
doi:10.1103/PhysRevD.94.063529
[arXiv:1606.05585 [astro-ph.CO]].

\bibitem{Hanany:2005bc}
A.~Hanany and K.~Hashimoto,
JHEP \textbf{06} (2005), 021
doi:10.1088/1126-6708/2005/06/021
[arXiv:hep-th/0501031 [hep-th]].



\bibitem{Mitridate:2023oar}
A.~Mitridate, D.~Wright, R.~von Eckardstein, T.~Schr\"oder, J.~Nay, K.~Olum, K.~Schmitz and T.~Trickle,
[arXiv:2306.16377 [hep-ph]].

\bibitem{Lamb:2023jls}
W.~G.~Lamb, S.~R.~Taylor and R.~van Haasteren,
Phys. Rev. D \textbf{108} (2023) no.10, 103019
doi:10.1103/PhysRevD.108.103019
[arXiv:2303.15442 [astro-ph.HE]].

\bibitem{Abbott:2021xxi}
R.~Abbott \textit{et al.} [KAGRA, Virgo and LIGO Scientific],
Phys. Rev. D \textbf{104} (2021) no.2, 022004
doi:10.1103/PhysRevD.104.022004
[arXiv:2101.12130 [gr-qc]].

\bibitem{Robson:2018ifk}
T.~Robson, N.~J.~Cornish and C.~Liu,
Class. Quant. Grav. \textbf{36} (2019) no.10, 105011
doi:10.1088/1361-6382/ab1101
[arXiv:1803.01944 [astro-ph.HE]].

\end{thebibliography}
\end{document}